 \documentclass[aps,preprint,superscriptaddress,groupedaddress]{revtex4}  
\usepackage{graphicx}  
\usepackage{dcolumn}   
\usepackage{bm}        
\usepackage{amssymb}   
\usepackage{amsmath}
\usepackage{dsfont}
\usepackage{color}

\newcommand{\be}{\begin{equation}}
\newcommand{\ee}{\end{equation}}
\newcommand{\bea}{\begin{eqnarray}}
\newcommand{\eea}{\end{eqnarray}}
\newcommand{\bee}{\begin{eqnarray*}}
\newcommand{\eee}{\end{eqnarray*}}

\newcommand{\nnu}{\nonumber\\}

\newcommand{\mt}{\tilde m}

\newcommand{\at}{\tilde a}

\newcommand{\Sig}{\bf\Sigma}

\def\mt{{\ifmmode\td M_t\else $\td M_t$\fi}}
\def\as{{\ifmmode\alpha_s\else$\alpha_s$\fi}}

\let\td=\tilde

\def\co#1{{\ifmmode{\cal O}_{#1}\else${\cal O}_{#1}$\fi}}
\def\cs#1{{\ifmmode{\cal S}_{#1}\else${\cal S}_{#1}$\fi}}
\def\at{{\ifmmode{\tilde A}\else$\tilde A$\fi}}

\def\fr#1.#2.{{#1\over #2}}

\def\mg{{\ifmmode M_{GUT}\else $M_{GUT}$\fi}}
\newcommand{\os}{\overline\Sigma}
\newcommand{\s}{\Sigma}
\newcommand{\Sigb}{{\overline\Sigma}}
\newcommand{\oot}{\overline {126}}

\newcommand{\ovl}{\overline}

\begin{document}




\title{ SO(10) grand unified theories with dynamical Yukawa couplings }
\author{ Charanjit S. Aulakh\footnote{aulakh@pu.ac.in; http://physics.puchd.ac.in/aulakh/}  }
\affiliation{Dept. of Physics, Panjab University\\
 Chandigarh, 160014, India}
\affiliation{Indian Institute of Science Education and Research
Mohali,\\ Sector 81, S. A. S. Nagar, Manauli PO 140306, India }
  \affiliation{Theory Division, CERN, Geneva \\
 CH-1213 Geneva 23, Switzerland}

 \author{  Charanjit K. Khosa}
 \affiliation{Dept. of Physics, Panjab University\\
 Chandigarh, 160014, India}

\date{\today}

\begin{abstract}
Renormalizable  SO(10) grand unified theories (GUTs), extended by $O(N_g)_F$ family gauge
symmetry, generate minimal supersymmetric Standard Model  flavour structure dynamically via
vacuum expectation values of ``Yukawon'' Higgs multiplets.
        For concrete illustration and  calculability, we   work with  the
 fully realistic  minimal  supersymmetric GUTs
  based on the $\bf{210 \oplus \oot\oplus
 126} $  GUT Higgs system - which were  already parameter counting
 minimal  relative to other realistic models. $SO(10)$ fermion
  Higgs channels  $\bf{\oot,10}$($\mathbf{120}$)
      extend to symmetric(antisymmetric) representations of $O(N_g)_F$,
while \textbf{210,126} are symmetric.
  $N_g=3$   dynamical Yukawa generation reduces the
  matter fermion Yukawas  from 15 to 3 (21 to 5)   without (with)
the $\bf{120}$ Higgs. Yukawon  GUTs are  thus ultraminimal in
parameter counting terms. Consistent symmetry breaking is ensured
by a  hidden sector   Bajc-Melfo(BM) superpotential with a pair of
symmetric $O(N_g)$ multiplets $\phi,S $, of which the latter's
singlet part $S_s$ breaks supersymmetry and the traceless part $\hat S
$  furnishes flat directions to cancel the $O(N_g)$ D-term
contributions of the visible sector. Novel dark matter
candidates linked to flavour symmetry arise from both the BM
sector and GUT sector minimal supersymmetric Standard Model singlet pseudo-Goldstones.   These
relics   may be viable light($< 50 $ GeV) cold dark matter as
reported by DAMA/LIBRA. In contrast to the new minimal supersymmetric SO(10) grand unified theory (NMSGUT) even  sterile
neutrinos can appear in certain branches of the flavour symmetry
breaking without the tuning of couplings.

\end{abstract}


 \maketitle

  \section{\label{sec:level1} Introduction}
Renormalizable SO(10) GUTs have an   array of virtues.
\textbf{16}-plet  spinors   enclose   Standard Model (SM) families
  with   right handed neutrinos required for seesaw
neutrino masses. Natural quark lepton unification is achieved with
realistic fermion mass relations based on just three allowed
fermion mass generating Higgs irreducible
representations (irreps) (in the renormalizable case)
: $\bf{10,120,\oot}$. Matter parity($(-)^{3(B-L) }$) is a part
 of  GUT gauge symmetry, and
using only vacuum expectation values (VEVs) with even values of
Baryon Number($B$)  - Lepton Number ($L$) ensures
R-parity   is conserved to low scales in the supersymmetric
case\cite{RparitySO10}. The  lightest supersymmetric particle
(LSP)  being stable is an  ideal weakly interacting massive particle dark matter candidate.
SO(10) GUTs based on the $\bf{210 \oplus \oot\oplus 126 }$ Higgs
system \cite{aulmoh,ckn} are parameter counting
minimal\cite{abmsv}. The version called the new minimal supersymmetric SO(10) grand unified theory (NMSGUT) with the
\textbf{120}-plet   fits\cite{blmdm,nmsgut,bstabhedge}  all
fermion masses and makes distinctive predictions  and is thus
falsifiable. Notably,  we found it requires large soft
$A-$terms\cite{nmsgut} in 2008 : well before Higgs discovery
promoted general acceptance of this possibility. It also predicts
a \emph{normal} s-hierarchy, Bino LSP  and can yield a light smuon
as the
  next to lightest supersymmetric particle(NLSP), thus promoting a Bino as a good dark matter candidate
and making a significant supersymmetric contribution to the muon
magnetic moment anomaly. Recently, we showed that careful attention
to the quantum communication between the UV completion and its
effective theory \emph{through the light Higgs portal} yields
natural and generic suppression of fast proton decay in supersymmetric(SUSY)
GUTs\cite{bstabhedge}.

The appealing  speculation  that the observed \emph{dimensionless}
fermion Yukawa couplings actually arise via VEVs of  `spurion'
fields has a long history\cite{spurion}.  In Ref.\cite{koide}, these
fields -appropriately called ``Yukawa-on''s-   carry
representations of
    $ U(N_g)^{6}$ family symmetry  (in a theory with six SM
flavours ($Q,L,u^c_L,d^c_L,e^c_L,\nu^c_L$) and $N_g$ generations),
but  \emph{not}   of the SM gauge group. Given that the main hint
for flavour unification   is   the convergence of third generation
Yukawas at GUT scales,   Yukawon models with   flavour symmetry
broken at the GUT scale are well motivated. Our
work\cite{aulmoh,abmsv,ag2,blmdm,nmsgut,bstabhedge} on realistic
renormalizable supersymmetric SO(10) GUTs - which encodes the
fermion hierarchy in the SO(10) matter fermion Yukawa matrices and
pays careful attention to how the minimal supersymmetric SM (MSSM) Higgs emerges from the
multiple MSSM type doublets present in the UV theory-also
naturally and minimally identifies the associated Higgs multiplets
as candidate Yukawon multiplets. It thus yields a novel mechanism
whereby  the fermion hierarchy could emerge from a flavour
symmetric and renormalizable GUT . Thus, in  our work ``Yukawons''
  \emph{also carry representations of the  gauge (SM/GUT)
dynamics}.  In previous work, typically   the  dimension 1
Yukawa-on $\mathcal{Y}$ in  the Higgs vertex made it
non-renormalizable   (${\cal{L}}= f^c \mathcal{Y} f H
/\Lambda_\mathcal{Y} +...$), where  the  unknown high scale
$\Lambda_\mathcal{Y}$ controls Yukawa-on dynamics.
 In this paper, we  work out   how minimal SO(10)
GUTs\cite{aulmoh,ckn,abmsv,nmsgut}  provide a gauged $O(N_g)$
family symmetry  route to ``Yukawonifcation'' : with the GUT and
family symmetry breaking at the same scale: obviating the need for
non-renormalizable interactions and
 any extraneous scale $\Lambda_\mathcal{Y}$. As we shall see, the
  consistency conditions for the maintainability of the   Higgs portal to the UV completion of the
MSSM play a central role in determining just how the peculiarly
lopsided and ``senseless'' fermion hierarchy  is produced from the
 flavour symmetric and grand unified  UV completion.
 In our
 \cite{aulmoh,abmsv,blmdm,nmsgut,bstabhedge}
minimal SUSY GUTs, we   eschew   invocation  of discrete symmetries
and insist only  upon following the logic of
  SO(10) gauge symmetry. This
insistence,  combined with careful attention to the implications
of the emergence of a light  MSSM Higgs pair from the    $2
N_g(N_g +1)$ pairs  in the $O(N_g)$ extended minimal supersymmetric GUT(MSGUT), leads to an
effectively unique extension of the SO(10) gauge group by a
$O(N_g)$ family gauge symmetry for the $N_g$ generation case and
the dynamical emergence of   fermion hierarchy and mixing.

The plan of this paper is as follows: In Section II we explain
the basic mechanism of generation of Flavor hierarchy by
enforcement of the masslessness of MSSM Higgs. In Section III we
discuss the spontaneous symmetry breaking of the GUT and flavor
symmetries. In Section IV we show how to calculate the flavor
hierarchies associated with two generation (toy model) and three
generation (realistic case) example solutions.In Section V we
discuss our results and the research program they define. In
Appendix A we give the explicit values of the VEVs for our example
solutions. In Appendix B we give the numerical values of the
Yukawa hierarchies for the three cases studied in Section IV along
with the complete superheavy spectra for each case.

\section{MSGUTs and the  YUMGUT proposal}

 The  minimal SUSY SO(10) model\cite{aulmoh,ckn} consists of
 an SM singlet containing  Higgs  irreps, $\bf{(210 (\Phi ) \oplus \oot (\os) \oplus
 126(\s)) }$, responsible for GUT symmetry breaking \emph{and} right handed neutrino masses
($M_{\nu^c}\sim <\os>$)  and $\bf{10(H),120(\Theta)}$ Higgs. The
$\mathbf{10,120}$ have SM doublets but no SM
  singlets and thus   do not participate in GUT scale spontaneous symmetry breaking (SSB).
 The $ \mathbf{{\oot (\os)}}$ also contributes to both neutrino and charged
 fermion  masses. Gauging just  an  $O(N_g)$
     subgroup of the   $U(N_g) $ symmetry of the fermion kinetic
  terms is workable   because the use of complex representations of a unitary family
  group introduces anomalies and requires doubling of the Higgs
  structure to cancel anomalies and to permit holomorphic  invariants
   to be formed for the superpotential.  Worse,
  unitary symmetry   enforces vanishing of half the
  emergent matter Yukawa couplings. $O(N_g)$ family symmetry suffers from none
  of these defects and gauging it ensures that no Goldstone bosons
  arise when it is spontaneously broken.
       $O(3)$ family symmetry with  tracefull  six-dimensional  irrep
      has also been considered for  non-renormalizable, non-GUT  Yukawa-on
   \cite{koide} models. However, our  model  is renormalizable and
   GUT based and thus   radically different in its construction and
   implications.

  The GUT superpotential   has \emph{exactly} the same form as
  the MSGUT (see Refs. \cite{ag1,ag2,bmsv,nmsgut} for comprehensive details)  :
  \bea W_{GUT}&=&\mathrm{Tr}( m \Phi^2 + \lambda \Phi^3 + M \Sigb .\s +\eta \Phi .\Sigb.
  \s)\nnu
  &+& \Phi.H.(\gamma \Sigma +\bar\gamma  . \Sigb) + { M_H}  H.H)\label{WGUT}\nnu
  W_{F}&=& \Psi_A .((h H) + (f \s) + (g\Theta) )_{AB}\Psi_B \label{WF}\eea

We have shown how the  ${\bf{120}}$-plet is  included in $W_F$ but
will here study only  MSGUTs (i.e with $\mathbf{10,\oot}$).
Inclusion of the \textbf{120}-plet does not affect GUT SSB. The
\emph{only} innovation in Higgs structure  is that all
   the MSGUT Higgs fields now carry symmetric
  representation of the $O(N_g)$  family symmetry :
  $\{\Phi,\Sigb,\s,H\}_{AB}; A,B=1,2..N_g$ (under which the
  matter $\mathbf{16}$-plets $\psi_A$ are vector $N_g$-plets).    Couplings $h,f,g$ are
  single complex numbers while the Yukawons
  carry symmetric ($\mathbf{H,\Sigb}$) and antisymmetric ($\mathbf{\Theta_{AB}}$)
   representations of   $O(3)$  : as required by the transposition
    property of   relevant  SO(10) invariants. For $N_g=3$,
   real fermion mass parameters come  down from 15
    ($Re[h_{AA}],f_{AB})$   to just
    3 ($Re[h],f$) without the \textbf{120}-plet
 (6 additional to just 2 with the ${\mathbf{120}}$-plet). Thus this type of renormalizable flavour unified GUTs can
    legitimately be called a Yukawon ultraminimal GUTs(\textbf{YUMGUT}s).

   An essential component of our proposal  is that  the single pair of
  light Higgs doublets ${\ovl{H}},H$ in
  the effective MSSM  arise  as  combinations of the manifold
  pairs of MSSM doublets  contained in the full set of fields.
   The consistency(also known as ``fine -tuning'') conditions for obtaining a light pair are
   here seen positively as the ``open sesame'' to peek through  the
    Higgs portal into the UV completion. This is possible
    precisely because the effective light pair is so deeply
    entangled in the innards of the UV theory in spite of being a
    bonafide member of the low-energy effective theory i.e the
    MSSM.  After   imposing  $ Det\mathbf{ \cal{H} } =0$  on the doublet
   mass matrix $\mathbf{ \cal{H} } $, one pair,
   ${\mathbf{\overline{H}}}[1,2,-1]\oplus \mathbf{H}[1,2,1]$,
  remains light.   Its Yukawa couplings are determined by the
  ``Higgs-fractions'' specified by the left and right null
  eigenvectors of   $\mathbf{\cal{H}}$ \cite{abmsv,ag2,nmsgut}.
    $\mathbf{\cal{H}}$ depends on the family symmetry
    breaking SM singlet VEVs $(p,a,\omega)_{AB}\in <\Phi_{AB}>,
    (\sigma,\bar\sigma)_{AB} \in <\s,\Sigb>_{AB}$.  Here, for
    simplicity,
          we  ignore the  ${\bf{120}}$-plet,
    although it   is known that this multiplet is required for
    fully realistic fermion masses when there is no family
    symmetry\cite{blmdm,nmsgut}. Then the four- dimensional $\mathbf{\cal{H}}$ of the
     MSGUT becomes $2 N_g(N_g+1)$ dimensional. If the  ${\bf{120}}$-plet
    were included (NMSGUT), $\cal{H}$ would have dimension $ N_g(3 N_g +1)$.

\begin{widetext}
\bea  \cal {H} &=&\begin{pmatrix} -M_H &\bar\gamma\sqrt{3}
\Omega(\omega-a)&-\gamma\sqrt{3} \Omega(\omega+a) &-\bar\gamma
\Omega(\bar\sigma)\cr
 \gamma\sqrt{3} \Omega(\omega-a)&-(2 M  + 4 \eta \Omega(a-\omega))
 & \varnothing_d &  -2 \eta \sqrt{3} \Omega(\bar\sigma) \cr
-\bar\gamma\sqrt{3} \Omega(\omega+a)&\varnothing_d &-(2 M + 4 \eta
\Omega(\omega+a))&\varnothing_d &\cr
  - \gamma \Omega( \sigma)  &-2 \eta \sqrt{3} \Omega( \sigma)&\varnothing_d &
 -2 m  + 6 \lambda \Omega(\omega-a)\cr
\end{pmatrix} \label{calH} \nnu \eea
\end{widetext} $\varnothing_d$ is the d dimensional null square matrix.  The rows  are labelled by the $N_g(N_g+1)/2$-tuples
(ordered and normalized, for a symmetric $\phi_{AB},
  A,B=1..N_g$,  as $\{\phi_{11},\phi_{22},...\phi_{N_gN_g},
 \sqrt{2}\phi_{12},\sqrt{2}\phi_{13},.....,\sqrt{2}\phi_{N_g-1,N_g}\}$)\hfil\break
  containing MSSM type ${\overline{H}}[1,2,-1]$ doublets from
  $\mathbf{10,126,\oot,210}$. The columns are labelled  by  ${ {H}}[1,2, 1]$ doublets
  in the order $\mathbf{10,\oot,126,210}$. The matrix function $\Omega(\phi)$ has the form determined by the
 symmetric invariant $Tr[<V> .{\overline{H}} . H ]$.  For
 $N_g=2$, it is \bea\Omega_2[V]= \begin{pmatrix} V_{11} &0&V_{12}/\sqrt{2}  \cr
 0 &V_{22}&V_{12}/\sqrt{2} \cr V_{12}/\sqrt{2} & V_{12}/\sqrt{2}
 &(V_{11}+V_{22})/2
  \cr \end{pmatrix}\label{Omega2} \eea
with labels $\{{\overline H}_{11},{\overline
H}_{22},\sqrt{2}{\overline H}_{12}\}\oplus \{{ H}_{11},{
H}_{22},\sqrt{2}{  H}_{12}\}$. Imposing $Det{\cal{H}}=0$, fermion
Yukawa couplings of  MSSM are extracted by generalizing the MSGUT
procedure\cite{abmsv,ag1,ag2,nmsgut}.  For $N_g=2$, we get Yukawa
couplings  \bea Y_u &=&
\begin{pmatrix} \hat h \hat V_1 + \hat f \hat V_4 & (\hat h \hat
V_3 + \hat f \hat V_6)/\sqrt{2} \cr (\hat h \hat V_3 + \hat f \hat
V_6)/\sqrt{2} & \hat h \hat V_2 + \hat f \hat V_5 \cr
\end{pmatrix} \nnu
 Y_d &=&\begin{pmatrix} \hat h \hat W_1 + \hat f \hat W_7 & (\hat h
\hat W_3 + \hat f \hat W_9)/\sqrt{2} \cr (\hat h \hat W_3 + \hat f
\hat W_9)/\sqrt{2} & \hat h \hat W_2 + \hat f \hat W_8 \cr
\end{pmatrix} \nnu \hat h &=& 2\sqrt{2} h \qquad ;\qquad \hat f=-4
i\sqrt{\frac{2}{3}} f
 \label{yukawas}\eea  The Yukawas $Y_\nu,Y_l$ for the neutrinos and charged
leptons are obtained from $Y_u,Y_d$ by the replacement $\hat f
\rightarrow -3 \hat f$. $\hat V,\hat W$ are the normalized right
and left null eigenvectors of the mass matrix
$\mathbf{\mathcal{H}}$ and contain\cite{ag1,ag2} the ``Higgs
fractions".  It is obvious that   $|h|>>|f|$ implies
$Y_u=Y_\nu,Y_d=Y_l$. Right-handed neutrino masses are proportional
to $f <\overline \Sigma>$, so this condition will elevate type I
(and depress type II) seesaw light neutrino masses. On the other
hand, the type II seesaw mechanism is more complicated now since
the $O(1,3,-2)$ $SU(2)_L$ triplet sub-multiplets of
$\mathbf{\oot}$ ( the VEVs of which generate type II seesaw masses) are now
symmetric multiplets of the flavour group. Although
the same kind of conflicting demands on the magnitude of the
coupling $f$ that put paid to the MSGUT's claim to fit neutrino
masses[5]may well operate here it must actually be verified in
detail. Our results here  even suggest that when light sterile neutrinos
are present the achieved neutrino masses may be much larger: to
our knowledge this possibility was not considered earlier in
MSGUTs. Since our objective in this paper is not the actual
fitting of the observed masses but only the introduction of
flavour-gauged MSGUTs and the way in which they generate MSSM
couplings, we postpone the computation of type II seesaw masses to
a sequel.  We will however give numerical illustrations of the
difficulty with type I.

\section{Spontaneous Symmetry Breaking in YUMGUTs}
To demonstrate the feasibility of our proposal for  dynamical
fermion flavour mixing and mass hierarchy from a family symmetric
Lagrangian,  we must
  show   the GUT scale spontaneous symmetry breaking    leads to Yukawas  like those
of  the observed fermion hierarchy. Of course such a program will
be complete only after finding values of the -flavour bland-  GUT
parameter set compatible with all other requirements such as gauge
coupling unification, proton stability and constraints on exotic
processes such as quark and lepton flavour violation, cosmology
etc. Without family symmetry this was achieved, apart from a
failure in matching  neutrino masses\cite{ag2,blmdm}, in the MSGUT
and, after prolonged investigations during 2006-2013, fully in the
NMSGUT \cite{nmsgut,bstabhedge}. We are now attempting to build
upon those successes by trying to generate also flavour structure
\emph{on the basis of the GUT SSB which is flavour blind}. When
carrying this out at the fully realistic level we will need to
search  the -flavour blind and thus much  reduced- parameter set
of the YUMGUT for sets of parameters that yield fully realistic
Yukawas. The
  Yukawonification  searches will  have much   larger  mass matrices than the (N)MSGUT case   but
  are also simpler due to  the reduced parameter space.  However, before embarking upon this mammoth and exacting
  project, we wish to show that even for generic values of parameters, the flavour structure
   that emerges can have the main features (hierarchy, small quark mixing, large lepton mixing) actually observed.
   Thus here we only indicate the operative structure,  and show
feasibility by finding GUT scale    SSB  solutions for generic
values of flavour bland parameters that yield a fermion hierarchy
qualitatively similar to that required in actuality. The search
for realistic parameter sets is a long-term objective that is
motivated and justified by the in-principle feasibility we
demonstrate here for both $N_g=2$
 and $N_g=3$.

Giving VEVs\cite{abmsv,ag1,ag2} to the SM singlets $p,a,\omega $
of the $(\mathbf{210},d_s[N_g])$-plet and the $\sigma\in
(\mathbf{126},d_s[N_g]),\bar\sigma\in (\mathbf{\oot},d_s[N_g])$,
where $d_s[N_g]=N_g(N_g+1)/2$, the superpotential relevant to GUT
SSB has just the MSGUT form, with sums over flavour group indices:
\bea W&=& \mathrm{Tr }[m(p^2+3 a^2+6 \omega^2)+ 2 \lambda (a^3+3 p
\omega^2 +6 a \omega^2)]\nonumber\\&&+\mathrm{Tr}[M \bar\sigma
\sigma
 + \eta (p+3a-6 \omega)\frac{(\bar \sigma \sigma  +  \sigma \bar \sigma)}{2}]
  \eea
In the MSGUT case, the entire SSB  problem reduces\cite{abmsv} (i.e
for $N_g=1$ in present notation) to a  cubic equation for a single
complex variable $x=-\omega$ (with just one free parameter $ \xi =
M\lambda/m\eta $).
  YUMGUTs proposed here generalize
  that solution  to the  family symmetric case.
    Visible sector  F-terms  give  : \be  2 m (p-a) -2
\lambda a^2 +2\lambda \omega^2  =0
\label{pa}  \ee
\be  2 m(p+ \omega ) +  \lambda (p+2 a + 3 \omega)\omega
  +  \lambda \omega (p+2 a + 3 \omega) =0  \label{pw}
 \ee 
\be  M \sigma+\eta (\chi \sigma +\sigma \chi)/2 =0\label{homosig} \ee
\be M \bar\sigma+\eta (\chi \bar\sigma +\bar \sigma \chi)/2  =0 \label{homosigbar}\ee
 \be  \bar\sigma\sigma  + \sigma \bar\sigma  = -{\frac{4}{\eta}}(m p +3 \lambda
 \omega^2) \equiv   F  \label{sgbsg}
  \ee
where $\chi\equiv (p+3 a -6 \omega) $. Homogenous equations
(\ref{homosig},\ref{homosigbar})  can be put in the   form
$\mathbf{ \Xi \cdot\hat\Sig =\Xi  \cdot \hat\os =0} $, where
$\mathbf{\hat\Sig,\hat\os} $ are $N_g(N_g+1)/2$ dimensional
vectors  : e.g. for $N_g=2, \hat\Sig
=\{\sigma_{11},\sigma_{22},\sigma_{12}\}$, the matrix
$\mathbf{\Xi}$ involves the combinations $\chi_A=\chi_{AA} +\frac{M}{\eta} $
: \bea \mathbf{\Xi}=
\begin{pmatrix}   \chi_1& 0&\chi_{12} \cr 0&\chi_{2}&\chi_{12} \cr
\chi_{12} &\chi_{12}& \chi_1+\chi_2
\end{pmatrix} \eea

Nontrivial solutions of Eq. (\ref{homosig}) and (\ref{homosigbar})
for $\sigma,\bar\sigma $ exist only if   $Det[\mathbf{\Xi}]=0$.
  Once Eqs. (\ref{pa})-(\ref{sgbsg}) are
(numerically) solved, there remain only the D-term conditions
  $D_{B-L}=0$ (the only
 nontrivial condition from the vanishing of  SO(10) D terms) and
$D^A=0$  from $O(N_g)$.    In the MSGUT $D_{B-L}=0$ requires
$|\sigma|=|\bar\sigma|$ and
   since  $Arg[\sigma]-Arg[\bar\sigma]$   can be
   gauged away by $U(1)_{B-L}$, the D-terms   supplement   F-terms
   to a nicety. We shall retain this convenience by only
   considering solutions with $\sigma_{AB}=\bar{\sigma}_{AB}$.

   At this point, an obstacle arises.  $O(N_g)$  D-terms
   vanish for vanishing  visible sector  F terms
     only for trivial(flavour diagonal)  solutions, since     different
     flavour
     charge sectors have no reason to cancel. So
   additional fields with   VEVs  free to cancel the contribution ${\bar{D}_X}^A$
    of the GUT sector to $O(N_g)$ D terms are needed.  The extra  F terms must  be
     sequestered from the GUT sector to preserve the MSGUT SSB.
      We propose\cite{BMsugry} Bajc-Melfo
    two field superpotentials \cite{ray,BM} ( of structure $W=S (\mu_B \phi  + \lambda_B \phi ^2)$ ).
    Their   potentials
    have   local minima breaking  supersymmetry ($<F_{S}> \neq 0$), which  leave  the VEV
    $<S>$ undetermined  ($\phi$ gets a VEV).
      Radiative corrections that determine $<S>$ then provide an alternative
      realization\cite{BM} of the Witten\cite{witten} model for high-scale symmetry
      breaking triggered by a low scale symmetry breaking. Determination of
      the flat direction VEV by
     N=1 \emph{supergravity } corrections instead of radiative
     effects  was  also shown to be effective long ago\cite{ovrab,hilo}.

 In Ref.\cite{BMsugry} we  show  that coupling  Bajc-Melfo fields to
  supergravity   resolves  the flat direction and    fixes the VEV  $<S_s>\sim M_p$.
  If   $S,\phi$ are \emph{also} traceful symmetric multiplets  $(S,\phi)_{AB}$ of  $O(N_g)$,  then  the
  traceless symmetric multiplet    $(\hat{S}_{AB} $)
    is  also undetermined at
  tree level  and free   to cancel $\bar{D}_X^A$. One gets a gravity-mediated scenario
  in which the hidden sector breaking involves breaking of family
  symmetry and supersymmetry. The additional  fields
  include moduli like fields ($\hat S_{AB}$), which can be light
  enough to serve as  light dark matter with mass less than 50
  GeV. Such masses are
  favoured \cite{damalibra} by experiments such as DAMA/LIBRA but
   were unobtainable in MSGUTs earlier. They are signals of the
   hidden (sector)   connection between supersymmetry and family symmetry breaking.
  Polonyi and moduli problems \cite{modulicon}  may be evaded because of the
   rich  Yukawa and gauge dynamics in the hidden sector.  We assume  results of
  Ref. \cite{BMsugry}, which were designed to ensure that the $O(N_g)$ terms
   can be cancelled and supersymmetry broken.

\section{\emph{Solutions with flavour generation}} For
$N_g>1$, one must consider the cases that arise
   separately.    For $N_g=2$ \bea && Det[\mathbf{\Xi}] =
(\chi_1+\chi_2)(\chi_{12}^2- \chi_1\chi_2)=0 \nnu \Rightarrow &&
\chi_1=-\chi_2 \quad  {OR} \quad \chi_{12}  = \pm\sqrt{
\chi_1\chi_2}\eea
  If $2 \times 2$ minors of $\mathbf{\Xi}$ also vanish then additionally
  $(\chi_1+\chi_2) \chi_1=(\chi_1+\chi_2) \chi_2=\chi_{12}^2$ and  $
  \chi_1\chi_2=0 $. Then  $ \chi_1=\chi_2=\chi_{12}=0 $ so that $ \mathbf{\Xi}\equiv
  0$. Thus, $Rank[\mathbf{\Xi}]<2  $ implies $Rank[\mathbf{\Xi}]=0$ so that none of
  the six $\sigma,\bar\sigma$ are determined. In the $Rank[\mathbf{\Xi}]=0$  case,
    one ultimately finds that the GUT spectra in sectors that
    include ($SO(10)/G_{123}$)  coset gauginos always include
large colour  and electroweak  non-singlet  pseudo-Goldstone
multiplets that would ruin unification. Therefore, we focus on the
 non degenerate cases in which  $Det[\mathbf{\Xi}]=0$ but
$Rank[\mathbf{\Xi}]=2$ for $N_g$=2 and $Rank[\mathbf{\Xi}]\leq 5$
for $N_g$=3. To show that our scenario is at least prima facie
consistent, in Appendix A  we give solutions of Eqs.(\ref{pa})-(\ref{sgbsg}) with zero $D_{B-L}$ for $N_g=2$, $N_g=3$
and the corresponding value of $ \bar{D}_X$ for a
 set of   YUMGUT flavour bland parameters (similar to those shown to be
 applicable in realistic fits in the NMSGUT \cite{bstabhedge}). We also give the complete GUT
 mass spectrum (in units of $m/\lambda$) so as to show that although there are
 cases  with  acceptable fermion hierarchies
 where there are light sterile neutrinos and  light MSSM singlets that can be interesting  dark matter
 candidates, these are  not plagued by
 MSSM non singlet pseudo-Goldstone  chiral  supermultiplets which would ruin
 the gauge coupling  unification so central to the whole NMSGUT
 project.  In this section, we explain some of the choices we made while navigating  the
  algebraic complexity of finding solutions of the coupled
 highly  non-linear equations (6)-(10). First, as always in MSGUTs\cite{abmsv}, it is convenient
to  work in units of $m/\lambda $: $\{P,W,A,\tilde \sigma, \tilde{\bar\sigma},\tilde\chi\}$=$\frac{\lambda}{m} \{p,\omega,a,\sigma,\bar\sigma,\chi\}$, $\tilde\chi_A=\tilde\chi_{AA}+\xi$ which removes most of the parameter
clutter in favour of just two parameter ratios ($\xi=\lambda
M/\eta m$ and $\lambda/\eta)$. \be
2\left(\frac{m}{\lambda}\right)^2(P-A-A^2+W^2)=0 \label{dl6} \ee
\be \left(\frac{m}{\lambda}\right)^2(P+W+(P+2A+3W)W+W(P+2A+3W))=0
 \label{dl7}\ee \be \left(\frac{m}{\lambda}\right)^2(\xi
\tilde\sigma+(\tilde\chi \tilde\sigma+\tilde\sigma
\tilde\chi)/2)=0  \label{dl8} \ee \be
\left(\frac{m}{\lambda}\right)^2(\xi \tilde
{\bar\sigma}+(\tilde\chi \tilde{\bar\sigma}+\tilde{\bar\sigma}
\tilde\chi)/2)=0 \label{dl9}\ee \be
\tilde{\bar\sigma}\tilde\sigma+\tilde\sigma\tilde{\bar\sigma}=-\frac{4
\lambda}{\eta}(P+3 W^2)=\frac{\lambda^2 F}{m^2} =\tilde{F}\label{dl10}  \ee
From now on, we will use these dimensionless equations  for SSB analysis.
\subsection{$N_g=2$, $Rank[\mathbf{\Xi}]= 2$}
As explained earlier,
 we will always take $\tilde\sigma=\tilde{\bar\sigma}$.  $Det[\mathbf{\Xi}] $ (same form as earlier but now in terms of dimensionless VEVs)
has two factors $(\tilde\chi_1+\tilde\chi_2)$ and
$(\tilde\chi_{12}^2-\tilde\chi_1 \tilde\chi_2)$, and one out of these
should vanish. From $\mathbf{ \Xi \cdot\hat\Sig =0} $, we can
calculate $\tilde\sigma_{11}$ and $\tilde\sigma_{22}$ in terms of
$\tilde\sigma_{12}$. \be
\tilde\sigma_{11}=-\frac{\tilde\chi_{12}}{\tilde\chi_1}
\tilde\sigma_{12} \quad ; \quad
\tilde\sigma_{22}=-\frac{\tilde\chi_{12}}{\tilde\chi_2}
\tilde\sigma_{12} \ee \be
Det[\tilde\sigma]=\frac{(\tilde\chi_{12}^2-\tilde\chi_1
\tilde\chi_2)}{\tilde\chi_1 \tilde\chi_2} \tilde\sigma_{12}^2 \ee
Notice the existence of one common factor in $Det[\mathbf{\Xi}] $
and $Det[\tilde\sigma]$. For a type I seesaw to operate for all
neutrino masses, we need an invertible $\mathbf{\oot }$ VEV since the
Majorana mass matrix of the right-handed neutrinos is proportional
to $<\overline\Sigma>$. Thus, we consider only  the branch
$(\tilde\chi_1+\tilde\chi_2)=0$ for vanishing $Det[\mathbf{\Xi}]$.
Then, $\tilde\sigma^2$ is diagonal. Equation \eqref{dl10} then
implies  \bea  \tilde\sigma_{11}^2= &=& \frac{\tilde F_{11}
\tilde\chi_{12}^2 }{2(\tilde\chi_{12}^2 +\tilde\chi_1^2)}\nnu
\tilde F_{11} &=&
 \tilde F_{22}\qquad ;\qquad \tilde F_{12}=0\label{Fcnstrt}\eea We begin by solving the three equations in Eq.(14),
which is linear in $P$ and parameter free,  for the three components
of $P$. Using these values $\tilde\chi_1=-\tilde\chi_2$ i.e \bea (P_{11}+3
A_{11} +W_{11})-(P_{22}+3 A_{22} +W_{22})+ 2\xi=0\eea is solved
for $A_{11}$ and then $\tilde F_{12}=0$ for $A_{12}$ and finally
$\tilde F_{11}-\tilde F_{22}=0$ for $A_{22}$.  Then, we solve the
remaining equations(13),  which are now expressed in terms of $W$
and $\xi$. We solve numerically for $W$ for  convenient $\xi$
using a minimization method.

With $P,A,W,\tilde\sigma$ in hand (see Appendix A for numerical values of
these  for sample sets of YUMGUT parameters),  we find the
$N_g(N_g+1)/2$ values of $M_H$ for which $Det{\cal{H}}=0$. For
\emph{each } of these possible values, we can   diagonalize
  MSSM  matter Yukawas  (\ref{yukawas}) and   the neutrino mass matrices
  to determine the fermion hierarchy. Of course,  the light  neutrino  masses will
depend not only on the electroweak VEV and $\tan\beta$ in the
effective MSSM but also on the masses of the right-handed
neutrinos  that depend on the Grand Unification
scale\cite{abmsv,ag1,ag2}. Here, we wish only to illustrate that
fermion hierarchies qualitatively similar to those observed in
reality   can arise for quite generic values of the YUMGUT
parameters.

  The fermion hierarchies  we obtain in the $N_g=2$  case for $f\sim
   10^{-1}$ and for $f \sim 10^{-4}$ are exhibited in Table \ref{table:ng=2} and
   \ref{table:ng=2fsmall}  in
Appendix B. We see that while $f\sim h$ generates acceptable quark
   and lepton mixing the neutrino masses and mass splitting
   squared tend to be much too small as is clear  from Table \ref{table:ng=2}.
   On the other hand-as seen in Table \ref{table:ng=2fsmall}-  when we lower $f$ we can raise the type I seesaw neutrino
   masses to an acceptable level but we then lose   acceptable
   Yukawa couplings obtained for $f\sim h$. To close the argument
   for $f\sim h$,  one needs to look also at the  operation of type
   II seesaw. However, note that in the MSGUT type II was always
   sub dominant to type I  because the mass of $O_{-}[1,3,-2]$ is
   generically of GUT scale whereas the largest right-handed
   neutrino masses allowed by type I seesaw are of order
   $10^{13- 14}$ GeV. Even though we now have a triplet of
   $O_{-}[1,3,-2]$ fields, except at special points, we may expect
   the same large masses to suppress the type II seesaw masses far below
   type I masses. Still, detailed study may reveal interesting
   special cases. Generically, however, the above arguments lead
   us to expect  that introduction of the \textbf{120}-plet and a special
   scenario involving  the lowering effect of threshold corrections on down
   type quark masses of the first two generations\cite{blmdm,nmsgut} is necessary
   for   phenomenological viability.

Thus, the main point of this numerical exercise with a toy model is
that \emph{without any attempt at optimizing these parameters, we
find ``small'' quark and ``large'' lepton mixing angles   and a
hierarchy between generations by factors of 10.} Such values in
the pivotal\cite{realcore} 23 sector  inspire confidence that
 an optimizing  \emph{search } for flavour-bland GUT parameters  may
  do much better.

A further issue that we must consider is the possibility that the
mass  spectrum of the 26 different MSSM multiplet types that occur in YUMGUTs based upon
the  MSGUT contains unacceptable pseudo-Goldstone
multiplets which would  ruin coupling unification. In Table
\ref{Ngeq2PGoldies}, we exhibit the spectrum in units of the
superheavy mass $m/\lambda \sim 10^{16} $ GeV. It is seen that
there
 is indeed a Pseudo-Goldstone of multiplet  type
 G(see Refs. \cite{ag1,ag2} for notation) with MSSM quantum numbers
 $[1,1,0]$. However, since this is a gauge singlet, it does not
 affect the evolution of the MSSM gauge couplings or their
 unification. Such particles are novel candidates for dark matter.

\subsection{$N_g=3$ SSB}
 The symmetry breaking equations for the realistic case  $N_g=3$   are considerably more complex  to solve but the patterns they yield  are
 again of the right type and moreover they introduce  a panoply of possibilities including  cases with light sterile neutrinos. To see how these arise, note that once
  again $\tilde\sigma$ and $\tilde{\bar\sigma}$ equations can be written as \be \mathbf{ \Xi \cdot\hat\Sig =\Xi  \cdot \hat\os =0} \label{sig3} \ee
Now
\bea \mathbf{\Xi}=
\begin{pmatrix}   \tilde\chi_1& 0 & 0&\tilde\chi_{12} &\tilde\chi_{13}& 0 \cr 0&\tilde\chi_{2}&0&\tilde\chi_{12}&0 & \tilde\chi_{23}\cr
0 & 0& \tilde\chi_{3}&0 &\tilde\chi_{13}& \tilde\chi_{23}\cr
\tilde\chi_{12} & \tilde\chi_{12}& 0&\tilde\chi_{1}+\tilde\chi_{2} &\tilde\chi_{23}& \tilde\chi_{13}\cr
\tilde\chi_{13} & 0& \tilde\chi_{13}&\tilde\chi_{23} &\tilde\chi_{1}+\tilde\chi_{3}& \tilde\chi_{12}\cr
0 & \tilde\chi_{23}&\tilde\chi_{23} &\tilde\chi_{13}&\tilde\chi_{12}& \tilde\chi_{2}+\tilde\chi_{3}\end{pmatrix} \eea
and $\hat\Sigma=\{\tilde\sigma_{11},\tilde\sigma_{22}, \tilde\sigma_{33}, \tilde\sigma_{12}, \tilde\sigma_{13}, \tilde\sigma_{23}\}$.
\bea Det[\mathbf{\Xi}]&=& (\tilde\chi_1 \tilde\chi_{2} \tilde\chi_{3}-\tilde\chi_1
   \tilde\chi_{23}^2-\tilde\chi_{12}^2 \tilde\chi_3+2
   \tilde\chi_{12} \tilde\chi_{13} \tilde\chi_{23}-\tilde\chi_{13}^2 \tilde\chi_{2})\nonumber\\&& (\tilde\chi_{1}^2 \tilde\chi_2+
   \tilde\chi_{1}^2 \tilde\chi_{3}-\tilde\chi_{1} \tilde\chi_{12}^2-\tilde\chi_{1} \tilde\chi_{13}^2+\tilde\chi_{1}
   \tilde\chi_{2}^2+2 \tilde\chi_{1} \tilde\chi_{2} \tilde\chi_{3}+\tilde\chi_{1} \tilde\chi_{3}^2\nonumber\\&&-\tilde\chi_{12}^2
   \tilde\chi_{2}-2 \tilde\chi_{12} \tilde\chi_{13} \tilde\chi_{23}-\tilde\chi_{13}^2 \tilde\chi_{3}+\tilde\chi_{2}^2
   \tilde\chi_{3}-\tilde\chi_{2} \tilde\chi_{23}^2+\tilde\chi_{2}
   \tilde\chi_{3}^2-\tilde\chi_{23}^2 \tilde\chi_{3})\label{detXi} \eea
For a nontrivial solution to exist, $Det[\mathbf{\Xi}]=0$ is
necessary. We first  discuss the non degenerate case
$Rank[\mathbf{\Xi}]= 5$.

\subsubsection{$Rank[\mathbf{\Xi}]= 5$ }

By Kramer's rule, we can determine five of the $\tilde\sigma $ variables  in
terms of one undetermined variable (say $\tilde\sigma_{23}$). From Eq.
\eqref{sig3} \be
\begin{pmatrix} \Xi_5 & v \cr v^T & \tilde\chi_{2}+\tilde\chi_{3}
\end{pmatrix} \begin{pmatrix} \hat{\sigma} \cr \tilde\sigma_{23}
\end{pmatrix} = 0 \qquad  \Rightarrow \qquad
\hat{\sigma}=-(\Xi_5^{-1}v) \tilde\sigma_{23}\ee Here, $\Xi_5$ and $v $
are the upper left $5 \times 5$ block and sixth column (less the 66
element)  of $\mathbf{\Xi}$ respectively. $\hat{\sigma}$$=$
$(\tilde\sigma_{11}, \tilde\sigma_{22}, \tilde\sigma_{33}, \tilde\sigma_{12},
\tilde\sigma_{13})$. We can construct $\tilde\sigma$ from
$\hat{v}=-(\Xi_5^{-1}v)$ : \be \tilde\sigma= \begin{pmatrix}
{\hat{v}}_1 & {\hat{v}}_4 & {\hat{v}}_5 \cr
{\hat{v}}_4 & {\hat{v}}_2 & 1 \cr
 {\hat{v}}_5 & 1 & {\hat{v}}_3\end{pmatrix} \tilde\sigma_{23} \label{sigexp}\ee
 Then
 \bea Det[\tilde\sigma]=&&  \frac{Det[\mathbf{\Xi}] N_5(\tilde\chi)}
 {D_5(\tilde\chi)}\nnu
N_5(\tilde\chi)=&& (\tilde\chi_{13}^2 \tilde\chi_2 - 2 \tilde\chi_{12} \tilde\chi_{13} \tilde\chi_{23}
+ \tilde\chi_1 \tilde\chi_{23}^2 + \tilde\chi_{12}^2 \tilde\chi_3 -
     \tilde\chi_1 \tilde\chi_2 \tilde\chi_3) (\tilde\chi_{12} \tilde\chi_{13}^2 + \tilde\chi_{13} \tilde\chi_2 \tilde\chi_{23} - \tilde\chi_1 \tilde\chi_{12} \tilde\chi_3 \nonumber\\&&-
     \tilde\chi_{12} \tilde\chi_2 \tilde\chi_3) (-\tilde\chi_{12}^2 \tilde\chi_{13} + \tilde\chi_1 \tilde\chi_{13} \tilde\chi_2 + \tilde\chi_{13} \tilde\chi_2 \tilde\chi_3 -
     \tilde\chi_{12} \tilde\chi_{23} \tilde\chi_3)\tilde\sigma_{23}^3  \nnu
D_5(\tilde\chi)=&&  (-\tilde\chi_1 \tilde\chi_{12}^2 \tilde\chi_{13}^2 + \tilde\chi_1^2
\tilde\chi_{13}^2 \tilde\chi_2 - \tilde\chi_{12}^2 \tilde\chi_{13}^2 \tilde\chi_2 +
    \tilde\chi_1 \tilde\chi_{13}^2 \tilde\chi_2^2 + \tilde\chi_1^2 \tilde\chi_{12}^2 \tilde\chi_3 \nonumber\\&& - \tilde\chi_{12}^2 \tilde\chi_{13}^2 \tilde\chi_3 - \tilde\chi_1^3 \tilde\chi_2 \tilde\chi_3 +
   \tilde\chi_1 \tilde\chi_{12}^2 \tilde\chi_2 \tilde\chi_3 + \tilde\chi_1 \tilde\chi_{13}^2 \tilde\chi_2 \tilde\chi_3 - \tilde\chi_1^2 \tilde\chi_2^2 \tilde\chi_3 + \tilde\chi_{13}^2 \tilde\chi_2^2 \tilde\chi_3 \nonumber\\&&-
   2 \tilde\chi_{12} \tilde\chi_{13} \tilde\chi_2 \tilde\chi_{23} \tilde\chi_3 + \tilde\chi_1 \tilde\chi_2 \tilde\chi_{23}^2 \tilde\chi_3 + \tilde\chi_1 \tilde\chi_{12}^2 \tilde\chi_3^2 -
   \tilde\chi_1^2 \tilde\chi_2 \tilde\chi_3^2 + \tilde\chi_{12}^2 \tilde\chi_2 \tilde\chi_3^2\nonumber\\&& - \tilde\chi_1 \tilde\chi_2^2 \tilde\chi_3^2  )^3  \eea
Thus \bea \because \quad Det[\tilde\sigma] \sim Det[\mathbf{\Xi}]
\Rightarrow Det[\tilde\sigma]=0=Det[M_{\nu^c}] \eea  Since the  Majorana
mass matrix $M_{\nu^c}$ for right handed neutrinos is  generated
by the VEV $\tilde{\bar\sigma}$, it follows that   the spectrum includes
one or more light sterile neutrino supermultiplets for which the fermionic
parts will get Dirac masses only. Thus, if there is one zero
eigenvalue (when the other factors in $Det[\tilde\sigma]$ are neither
zero nor singular), one will have two superheavy  and one zero
Majorana mass for the right-handed neutrinos. The two superheavy
ones should be integrated out to give a $4 \times 4$ mass matrix
for the light neutrinos. Similarly, in the case of two zero
eigenvalues, one would integrate out just one right-handed
neutrino(see below). Of course the mixing structure is then more
complicated and must be studied case by case. These are still very
interesting cases phenomenologically and it is remarkable that
besides the standard three light Majorana neutrinos scenario typical
to the NMSGUT, YUMGUTs also throw up scenarios with mixed
Majorana-Dirac masses.

 One   proceeds to solve the $ Det[\mathbf{\Xi}]=0$ condition. The
prima facie simpler possibility is \bea  \tilde\chi_1 =
\frac{(\tilde\chi_{13}^2 \tilde\chi_2 -
  2 \tilde\chi_{12} \tilde\chi_{13} \tilde\chi_{23} + \tilde\chi_{12}^2 \tilde\chi_3)}{(\tilde\chi_2
\tilde\chi_3)-\tilde\chi_{23}^2  }  \eea We know $\tilde\sigma's$
in terms of $\tilde\sigma_{23}$ (Eq.\eqref{sigexp}). As in
the $N_g$=2 case, we calculated $\tilde\sigma_{23}$ from one of the
equations of Eq.(\ref{dl10}) and $P$ completely  using
Eq.(\ref{dl7}). We then solved remaining equations for $A, W$ using
a search program.

In the example solution given in the Appendix A, the $\tilde\sigma$
VEV has two zero eigenvalues because the  $ Det[\mathbf{\Xi}]$
factor that vanishes
\[(\tilde\chi_1 \tilde\chi_{2} \tilde\chi_{3}-\tilde\chi_1
   \tilde\chi_{23}^2-\tilde\chi_{12}^2 \tilde\chi_3+2
   \tilde\chi_{12} \tilde\chi_{13} \tilde\chi_{23}-\tilde\chi_{13}^2 \tilde\chi_{2})\]
actually  occurs twice in $Det[\tilde\sigma]$. We need to integrate out
one heavy right handed neutrino \be W_{lep}={\bar \nu}^T_{A}
Y^\nu_{AB}\nu_B+\frac{1}{2}{\bar
\nu}^T_{A}M^{\bar{\nu}}_{AB}{\bar\nu}_{B} \ee where $A,B={1,2,3}$
and $a,b$=1,2.
 \be
 {\bar\nu}_{3}=-\frac{Y^\nu_{3A}\nu_A}{M_{33}^{\bar{\nu}}} \ee
Effective superpotential is given
 by (in a diagonal right-handed neutrino basis)
 \be W_{eff}={\bar\nu}^T_{a}
Y^\nu_{aB}\nu_B+\frac{1}{2}{\bar\nu}^T_{a}M_{aa}{\bar{\nu}}_{a}
-\nu_A^T\biggr(\frac{1}{2}\frac{Y^\nu_{3A} Y^\nu_{3B}
 }{M_{33}^{\bar{\nu}}}\biggr)\nu_B
\ee Reading off the light neutrino mass matrix components, we get
\be \kappa_{AB}=-  (Y^{ \nu})^T_{A3}M_{33}^{-1}(Y^{ \nu})_{3B} \ee
\be  M_{light} =\frac{1}{2}
\begin{pmatrix} \kappa_{11} & \kappa_{12} & \kappa_{13} &
Y^{\nu}_{11}& Y^{\nu}_{21}\cr \kappa_{21} & \kappa_{22} &
\kappa_{23} & Y^{\nu}_{12}& Y^{\nu}_{22} \cr \kappa_{31} &
\kappa_{32} & \kappa_{33} & Y^{\nu}_{13}& Y^{\nu}_{23} \cr
Y^{\nu}_{11}& Y^{\nu}_{12} &Y^{\nu}_{13} & 0& 0\cr Y^{\nu}_{21}&
Y^{\nu}_{22} &Y^{\nu}_{23} & 0& 0 \end{pmatrix} \ee

 The doublet  Higgs mass matrix becomes 24 dimensional
and can be written using $\Omega_3[V]$ which can be determined as
$\Omega_2$ was in Eq. (\ref{Omega2}). One gets \bea
\Omega_3[V] =
\begin{pmatrix} V_{11} & 0 & 0 & \frac{V_{12}}{\sqrt{2}} & \frac{V_{13}}{\sqrt{2}} & 0 \cr
0 & V_{22} & 0 & \frac{V_{12}}{\sqrt{2}} & 0 & \frac{V_{23}}{\sqrt{2}}\cr
0& 0 & V_{33} & 0 &\frac{V_{13}}{\sqrt{2}} & \frac{V_{23}}{\sqrt{2}}\cr
\frac{V_{12}}{\sqrt{2}}&\frac{V_{12}}{\sqrt{2}}&0 &\frac{V_{11}+V_{22}}{2}& \frac{V_{23}}{2} &\frac{V_{13}}{2} \cr
\frac{V_{13}}{\sqrt{2}} & 0 & \frac{V_{13}}{\sqrt{2}}& \frac{V_{23}}{2}&\frac{V_{11}+V_{33}}{2}&\frac{V_{12}}{2} \cr
 0 &\frac{V_{23}}{\sqrt{2}} & \frac{V_{23}}{\sqrt{2}}& \frac{V_{13}}{2}&\frac{V_{12}}{2}&\frac{V_{22}+V_{33}}{2}
\end{pmatrix} \eea

The formulas for the Yukawas are now  simple extensions of the
$N_g$=2 case : \bea Y_u &=&
\begin{pmatrix} \hat h \hat V_1 + \hat f \hat V_7 & (\hat h \hat
V_4 + \hat f \hat V_{10})/\sqrt{2} & (\hat h \hat V_5 + \hat f \hat V_{11})/\sqrt{2}  \cr
(\hat h \hat V_4 + \hat f \hat V_{10})/\sqrt{2} & \hat h \hat V_2 + \hat f \hat V_8 & (\hat h \hat V_6 + \hat f \hat V_{12})/\sqrt{2}\cr
(\hat h \hat V_5 + \hat f \hat V_{11})/\sqrt{2} & (\hat h \hat V_6 + \hat f \hat V_{12})/\sqrt{2} &  \hat h \hat V_3 + \hat f \hat V_9
\end{pmatrix} \nnu
 Y_d &=&\begin{pmatrix} \hat h \hat W_1 + \hat f \hat W_{13} & (\hat h
\hat W_4 + \hat f \hat W_{16})/\sqrt{2}  & (\hat h
\hat W_5 + \hat f \hat W_{17})/\sqrt{2} \cr (\hat h \hat W_4 + \hat f
\hat W_{16})/\sqrt{2} & \hat h \hat W_2 + \hat f \hat W_{14} & (\hat h \hat W_6 + \hat f
\hat W_{18})/\sqrt{2} \cr  (\hat h \hat W_5 + \hat f
\hat W_{17})/\sqrt{2} & (\hat h \hat W_6 + \hat f
\hat W_{18})/\sqrt{2} & \hat h \hat W_3 + \hat f \hat W_{15}
\end{pmatrix}
 \label{yukawas3d}\eea

The Yukawa couplings, mixing angles and neutrino masses for an
example  solution in this case are exhibited in Table
\ref{table:rank5}.  It is interesting that in this case, four of
the  neutrino masses are in the desirable  range of $10^{1\div 2}$
meV even though $Y_u\neq Y_\nu, Y_d\neq Y_l $. Effectively, the two vanishing $M_{\nu^c}$ eigenvalues
 combined with the mixing of $\nu$ and $\nu^c$ via Yukawa couplings
   gives five light neutrinos of which at least two have
significant light sterile  admixture. However, such  non-minimal scenarios are not yet
firmly excluded by the data. The mixing angles can be extracted
much as in the case without the neutrino, but the phenomenology is
more subtle. The admixture of light sterile neutrinos
    to obtain neutrino masses in the observed (meV) range
    indicates a possible route-not eliminated
in Refs. \cite{blmdm,bertschwmal}- also for the MSGUT by which the no
go proved for neutrino masses in the MSGUT  might be evaded. In
Tables \ref{tracefullrank5spec} and \ref{htrank5-6dim}, we exhibit the
GUT spectra including three singlet pseudo-Goldstones.

\subsubsection{$Rank[\mathbf{\Xi}]= 4$}
In this case, all the $5 \times 5 $ minors of $\mathbf{\Xi}$ should also vanish along
 with $Det[\mathbf{\Xi}]$. One can determine four $\tilde\sigma's$ in terms of the remaining two.
 Suppose we calculate $\tilde\sigma_{11}$, $\tilde\sigma_{22}$, $\tilde\sigma_{33}$ and $\tilde\sigma_{12}$ in
 terms of ($\tilde\sigma_{13}$, $\tilde\sigma_{23}$). Then \be \tilde\sigma= A \tilde\sigma_{13}+ B \tilde\sigma_{23} \ee
   where the matrices A and B   are functions of the  $\tilde\chi_{AB}$. Now, $Det[\tilde\sigma]$ factors will involve
 $\tilde\sigma_{13}$, $\tilde\sigma_{23}$ instead of involving purely $\tilde\chi$ elements with the undetermined
 $\tilde\sigma$ component
  factorizing  outside as in case of $Rank[\mathbf{\Xi}]= 5$. Even if one assumes
  $\tilde\sigma_{13}=\tilde\sigma_{23}$ (although there is no reason to assume this),
  we find that $Det[\tilde\sigma]$ still has no common factor with $Det[\mathbf{\Xi}]$. So
  $Rank[\mathbf{\Xi}]=4$  could be a workable scenario with type I seesaw neutrino masses and without light sterile neutrinos.
   Now, we discuss the conditions for vanishing minors.

 All the $5 \times 5 $ minors have one common factor. Ideally, one can solve the
  equations by putting  the common factor of the minors equal to zero  but  the resulting system becomes  very complicated.
   For simplicity, we equate  to zero  two factors from the dimension-5 minors the
vanishing of which implies the vanishing of
   all the dimension-5 minors.
   In all,  we get three conditions, one from $Det[\mathbf{\Xi}]$=0 (Eq. \ref{detXi} )and two from $5 \times 5 $ minors.

 Minors of order 5 are null if we
require \bea  \tilde\chi_{13}\tilde\chi_{23}-\tilde\chi_{12} \tilde\chi_{3}=0 \nnu
  \tilde\chi_{12}^2 \tilde\chi_{13} + \tilde\chi_{12} \tilde\chi_{2} \tilde\chi_{23} + \tilde\chi_{13}
  \tilde\chi_{23}^2=0\eea
 We solve the above equations for  $\tilde\chi_2$ and $\tilde\chi_3$ :
\bea \tilde\chi_1 =\frac{\tilde\chi_{12}\tilde\chi_{13}}{\tilde\chi_{23}} \quad ; \quad
\tilde\chi_2
=\frac{-\tilde\chi_{13}(\tilde\chi_{12}^2+\tilde\chi_{23}^2)}{\tilde\chi_{12}\tilde\chi_{23}}
\quad ; \quad \tilde\chi_3 =\frac{\tilde\chi_{13}\tilde\chi_{23}}{\tilde\chi_{12}}  \eea

Now we  have four $\tilde\sigma$ equations along with above three conditions.
We used the third extra condition to fix the $\xi$ parameter to
maintain the consistency of the system of equations. Therefore,
$\xi$ is not a random parameter but is fixed in terms of
VEVs$(P,A,W)$. The alternatives to this will be examined
elsewhere.

We calculate the  eigenvalues of SM fermions and
neutrino Yukawas (Eq. \eqref{yukawas3d}) along with quark and lepton mixing angles for
large and small values of $f$  and a random illustrative set of
superpotential parameters. The results are given as Tables \ref{table:tracefulldata} and
\ref{table:tracefulldata-fsmall}.

As in the $N_g=2$ case,  observe that if $f$ affects the charged
fermion Yukawas significantly, avoiding $Y_u=Y_\nu, Y_d=Y_l$, and
generates appreciable mixing  then the type I seesaw neutrino
masses  are much too small.  While if we boost the type I seesaw
masses by lowering $f$, $Y_\nu\simeq Y_u,Y_l\simeq Y_d$ and Cabibbo Kobayashi Maskawa(CKM)
angles are negligible. The complete superheavy spectra are given
in Tables \ref{tracefullspec} and \ref{htrank4-6dim},  and again we
have (4) standard model singlet pseudo-Goldstones.
\subsection{ Traceless Symmetric multiplets of O(3)}
As we saw earlier, the (reducible) six-dimensional symmetric representation
of O(3) with equal superpotential couplings for traceless(5-plet)
and singlet part  led, in the non-degenerate case
($Rank[\mathbf{\Xi}]=5$)  to a light sterile neutrino(s). Another
alternative is to   utilize only traceless representations in the
(visible sector) GUT so that
$P,A,W,\tilde\sigma,\tilde{\bar\sigma}$ are all traceless
symmetric matrices. This reduces the dimension and rank  of the
homogeneous system (${\bf {\Xi. \hat \Sigma=0}}$) and makes it
possible to find non-singular $\tilde\sigma(\tilde{\bar\sigma})$
VEVs thus avoiding light sterile neutrinos. We write the traceless
symmetric $3 \times 3 $ representation as
\[ \hat \phi_{AB}= \phi_{11} \frac{(\lambda_3)_{AB}}{\sqrt{2}} +\phi_{22} \frac{(\lambda_8)_{AB}}
{\sqrt{2}}+ \frac{\phi_{KL}}{\sqrt{2}} \delta^K_{(A} \delta^L_{B)}
\] Here, $\lambda_3$ and $\lambda_8$ are the usual  diagonal $3\times 3$ Gell-Mann
matrices. Matrix $\mathbf{\Xi}$ is given by

\be \mathbf{\Xi}=\begin{pmatrix} -\tilde\chi_{22} + 2 \xi & -\tilde\chi_{11} - \tilde\chi_{22} + \xi & \tilde\chi_{12} &
  0 & -\tilde\chi_{23} \cr -\tilde\chi_{11} - \tilde\chi_{22} + \xi & -\tilde\chi_{11} + 2 \xi & \tilde\chi_{12} & -\tilde\chi_{13} &
  0 \cr \frac{\tilde\chi_{12}}{2} & \frac{\tilde\chi_{12}}{2} & \frac{(\tilde\chi_{11} + \tilde\chi_{22} + 2 \xi)}{2} & \frac{\tilde\chi_{23}}{2} & \frac{\tilde\chi_{13}}{
  2} \cr 0 & -\frac{\tilde\chi_{13}}{2}& \frac{\tilde\chi_{23}}{2} &  \frac{-\tilde\chi_{22} + 2 \xi}{2}& \frac{\tilde\chi_{12}}{2} \cr -\frac{\tilde\chi_{23}}{2} &
   0 & \frac{\tilde\chi_{13}}{2} & \frac{\tilde\chi_{12}}{2} &  \frac{-\tilde\chi_{11} + 2 \xi}{2} \end{pmatrix} \ee
$\hat\Sigma=\{\tilde\sigma_{11},\tilde\sigma_{22},
\tilde\sigma_{12}, \tilde\sigma_{13}, \tilde\sigma_{23}\}$. As
discussed above it is useful to define a  matrix
function($\Omega[V]$ ) to write the mass matrices in
compact notation and is specified by the symmetric invariant for
the singlet  product of three traceless symmetric irreps. In the
present scenario, it has the following form : \be \Omega'_3[V] =
\begin{pmatrix} \frac{V_{11}+V_{22}}{2} & \frac{V_{11}-V_{22}}{2 \sqrt{3}} & 0 & \frac{V_{13}}{2} & -\frac{V_{23}}{2} \cr
\frac{V_{11}-V_{22}}{2 \sqrt{3}} & \frac{-V_{11}-V_{22}}{2} & \frac{V_{12}}{\sqrt{3}} & -\frac{V_{13}}{2\sqrt{3}} & -\frac{V_{23}}{2\sqrt{3}}\cr
0& \frac{V_{12}}{\sqrt{3}} & \frac{V_{11}+V_{22}}{2} &\frac{V_{23}}{2} & \frac{V_{13}}{2}\cr
\frac{V_{13}}{2}&-\frac{V_{13}}{2\sqrt{3}} &\frac{V_{23}}{2}& -\frac{V_{22}}{2} &\frac{V_{12}}{2} \cr
-\frac{V_{23}}{2} & -\frac{V_{23}}{2\sqrt{3}}&\frac{V_{13}}{2}&\frac{V_{12}}{2}&-\frac{V_{11}}{2}
\end{pmatrix}
 \ee
Rows and columns of the Higgs mass matrix are labelled by $\{\frac{(\bar
H_{11}-\bar H _{22})}{\sqrt{2}}, \sqrt{\frac{3}{2}}({\bar
H}_{11}+{\bar H}_{22}),\sqrt{2} \bar H_{12},\sqrt{2} \bar
H_{13},\sqrt{2} \bar H_{23}\}$ and $\{\frac{( H_{11}- H
_{22})}{\sqrt{2}}, \sqrt{\frac{3}{2}}({ H}_{11}+{
H}_{22}),\sqrt{2}  H_{12},\sqrt{2}  H_{13},\sqrt{2}  H_{23}\}$.
   Up and down quark Yukawas are given as :
{\small{\bea Y_u &=&
\begin{pmatrix} \hat h (\frac{\hat V_1}{\sqrt{2}}+\frac{\hat V_2}{\sqrt{6}} )+ \hat f( \frac{\hat V_6}{\sqrt{2}}+\frac{\hat V_7}{\sqrt{6}}) &
 \hat h \frac{\hat
V_3}{\sqrt{2}} + \hat f \frac{\hat
V_8}{\sqrt{2}} &  \hat h \frac{\hat
V_4}{\sqrt{2}} + \hat f \frac{\hat
V_9}{\sqrt{2}}   \cr \hat h \frac{\hat
V_3}{\sqrt{2}} + \hat f \frac{\hat
V_8}{\sqrt{2}} & \hat h (-\frac{\hat V_1}{\sqrt{2}}+\frac{\hat V_2}{\sqrt{6}} )+ \hat f( -\frac{\hat V_6}{\sqrt{2}}+\frac{\hat V_7}{\sqrt{6}}) &\hat h \frac{\hat
V_5}{\sqrt{2}} + \hat f \frac{\hat
V_{10}}{\sqrt{2}}  \cr \hat h \frac{\hat
V_4}{\sqrt{2}} + \hat f \frac{\hat
V_9}{\sqrt{2}}    & \hat h \frac{\hat
V_5}{\sqrt{2}} + \hat f \frac{\hat
V_{10}}{\sqrt{2}} & -2 \hat h \frac{ \hat V_2}{\sqrt{6}} -2 \hat f \frac{ \hat V_7 }{\sqrt{6}}
\end{pmatrix} \nnu
Y_d &=& \begin{pmatrix} \hat h (\frac{\hat W_1}{\sqrt{2}}+\frac{\hat W_2}{\sqrt{6}} )+ \hat f( \frac{\hat W_{11}}{\sqrt{2}}+\frac{\hat W_{12}}{\sqrt{6}}) &
 \hat h \frac{\hat
W_3}{\sqrt{2}} + \hat f \frac{\hat
W_{13}}{\sqrt{2}} &  \hat h \frac{\hat
W_4}{\sqrt{2}} + \hat f \frac{\hat
W_{14}}{\sqrt{2}}   \cr \hat h \frac{\hat
W_3}{\sqrt{2}} + \hat f \frac{\hat
W_{13}}{\sqrt{2}} & \hat h (-\frac{\hat W_1}{\sqrt{2}}+\frac{\hat W_2}{\sqrt{6}} )+
 \hat f( -\frac{\hat W_{11}}{\sqrt{2}}+\frac{\hat W_{12}}{\sqrt{6}}) &\hat h \frac{\hat
W_5}{\sqrt{2}} + \hat f \frac{\hat
W_{15}}{\sqrt{2}}  \cr \hat h \frac{\hat
W_4}{\sqrt{2}} + \hat f \frac{\hat
W_{14}}{\sqrt{2}}    & \hat h \frac{\hat
W_5}{\sqrt{2}} + \hat f \frac{\hat
W_{15}}{\sqrt{2}} & -2 \hat h \frac{ \hat W_2}{\sqrt{6}} -2 \hat f \frac{ \hat W_{12} }{\sqrt{6}}\end{pmatrix}
 \label{5dimyuk}\eea}}
 We
have solved the least degenerate ($Rank[\mathbf{\Xi}]=$ 4) case.
Yukawa  eigenvalues and mixing angles are given in Tables
\ref{table:tracelessdata}.

The complete mass spectra are given in Tables
\ref{tracelessspec} and \ref{htrank4-5dim}. There are three MSSM
\emph{singlet}
   pseudo-Goldstone supermultiplets left behind as a relic of
   the GUT and flavour symmetry breaking.

 From Tables \ref{table:tracefulldata} and \ref{table:tracelessdata}, we   conclude
  that it is possible to fit the  charged fermion and neutrino mixing data in the case without the \textbf{120}-plet
   without light sterile neutrinos but that the neutrino masses   are too small. Raising the neutrino masses requires that an additional
   multiplet  is required to contribute to fit  the charged fermion masses  and mixing :
   the obvious choice if duplication of Higgs is to be avoided is the  \textbf{120}-plet as in the case of the NMSGUT.
If we allow light sterile neutrinos (as we saw for $N_g=3, Rank[\mathbf{\Xi}]=5$)
then we can even get adequate neutrino masses without
$Y_u=Y_\nu,Y_d=Y_l$ and possibly viable phenomenology. This needs
to be studied.

\section{Discussion}
In this paper, we have proposed a gauge  unification of flavour in
tandem with the realistic grand unification achieved in the
context of the
NMSGUT\cite{abmsv,ag2,blmdm,nmsgut,bstabhedge}. The flavour
symmetry is broken along with SO(10) gauge symmetry at the GUT
scale by the same multiplets that break GUT symmetry. However, a
special (Bajc-Melfo) hidden sector was introduced\cite{BM}  to
allow supersymmetry preservation by the VEV of the flavour group
D-terms. The central idea exploited for generating flavour
structure  in the effective theory is the non-linear   percolation
of  flavour breaking   effects through the  imposition of  the
phenomenological constraint that a single pair of light doublets
emerges  in the effective MSSM from the multiple doublets in the
GUT. The consistent treatment of the implications of this
constraint had already paid\cite{bstabhedge} rich dividends such
as a solution of the perennial problem of fast dimension 5
operator proton decay in SUSY GUTs.  This resolution has driven
home the message that careful attention to the existence
conditions for the light Higgs   opens a portal into the entrails
of complicated UV completions precisely because  light MSSM Higgs
 arise as mixtures of doublets from multiple sources in the
GUT Higgs set and are thus connected to the entire UV completion
by non-gauge interactions  : in contrast to the rest of the fields
of the MSSM. In view of the understanding conferred, we call such
flavour ``tasty flavour'' and these models Yukawon ultraminimal
GUTs. This mechanism, like that\cite{bstabhedge}
  for suppressing $d=5$ B violation in the NMSGUT,
  depends on enforcing  consistency
conditions  between a  single light pair of MSSM doublets and the
multiple   doublets  in the full theory.
   A common apprehension is that  pseudo-Goldstone multiplets
   may arise due to the duplication  of Higgs multiplets under family symmetry.
   However, we have checked  the
\emph{complete} spectra in  the GUT sector and only present those
solutions without pseudo-Goldstone multiplets that ruin
unification (which do occur in certain degenerate cases). The
values of the fields and complete associated spectra for the
example solutions are given in Appendix B. This   shows that the
appearance of pseudo-Goldstones is \emph{not} inevitable except
the SM singlet sector which does not effect unification and
moreover furnishes dark matter candidates or otherwise constrains
the flavour symmetry introduced.

Therefore, we expect that there should  exist regions of the (much
reduced relative to MSGUTs)  flavour bland  parameter space of
YUMGUTs that generate actual MSSM Yukawas. Determination of GUT
parameter values that generate the observed set of Yukawa
couplings, symmetry breaking, neutrino masses, B-decay and other
exotic phenomenon rates requires an elaborate investigation  using
a generalization of   computer  codes   by which
 we obtained  completely  realistic fits of all fermion data and
distinctive    predictions  for SUSY spectra in the context of the
NMSGUT\cite{nmsgut}.  This will be reported in sequels.  In this
paper, we have shown how even random values of the YUMGUT
parameters can generate  a fermion hierarchy at least
qualitatively similar to the one actually observed. An important
point that emerges   is that an effective theory with light
sterile neutrinos can emerge naturally from the spontaneous
symmetry breaking so that even scenarios with relatively large
values of $f$ may lead to a viable phenomenology. This discovery
underlines the need to reinvestigate the MSGUT to explore the
possibility that it may still yield   acceptable neutrino masses
by giving ultralight  sterile neutrinos mixed through Yukawa
couplings with active neutrinos( which are therefore heavy enough)
even when the $\mathbf{\oot}$ coupling is large enough to
contribute significantly to the charged fermion masses and thus
type I masses are negligible. This possibility had been overlooked
earlier\cite{blmdm,bmsvdecem,bertschwmal}. Of course, the other
possibility is that an acceptable low-energy theory arises by
mimicking the NMSGUT fit i.e $\mathbf{10,120}$ couplings
responsible for charged fermion masses and small $f,
<\mathbf{\oot}>$  give   right-handed neutrinos light enough to
give acceptable type I seesaw masses for three light left-handed
neutrinos and no sterile neutrinos.

A striking feature of the model is the ubiquity of MSSM singlet
pseudo-Goldstones from the GUT symmetry breaking which occur in
each one of the models studied. In some sense, these are ``moduli''
multiplets familiar in other contexts, such as string theory,
which generate families. They may  serve as a welcome signal of,
and constraint on, these models. Soft SUSY breaking and radiative
corrections to their masses and the associated phenomenology and
cosmological implications deserve detailed  investigation.

 To summarize, R-parity preserving minimal SUSY SO(10) theories offer a
 radical conceptual simplification that melds the problems of the
 fermion flavour  hierarchy and GUT symmetry breaking in a  novel way.
  Innovation arises from a very topical emphasis on  how the Higgs
  portal opens into the UV completion of the MSSM.
    Our model  for family  and  GUT unification is based on the
conviction that, as once promised by string theory, it is the
number of couplings, not the number of fields,  that should be
minimized.   SO(10) ``Yukawonification" is   concrete,
 calculable (at least for (N)MSGUTs where much of the tedious
 but directly useable (see e.g.  Eq.(\ref{calH})) preparatory work
  on group theory\cite{ag1} and mass spectra \cite{ag1,ag2,bmsv,nmsgut,fuku04}
  is already  done) and falsifiable.
 Since $O(N_g)$ contains most of the commonly fancied discrete
 groups,  even   discrete group model builders  may benefit
 from the structural hints provided by $SO(10)\times O(N_g)$ models  along with a welcome
 calculability as regards the  ``vacuum alignment" :  specified here by
  parameters that determine
   both the  GUT SSB and MSSM mass-mixing data.
\eject
\section*{\bf {APPENDIX A : YUMGUT SUPERHEAVY VEVs }}
In this Appendix we give  solutions of the spontaneous symmetry
breaking in the toy ($N_g=2$)  and  realistic ($N_g=3$)  cases.

\subsection{$N_g$=2, $Rank[\mathbf{\Xi}]=$2}
The values of the VEVs of the
YUMGUT Higgs fields responsible for breaking $SO(10)\rightarrow
\rm{MSSM} $ in units of $m/\lambda\sim 10^{16}$ GeV are  \bea
 W&=& \begin{pmatrix}  0.141 - 0.203 i &  0.3168 + 0.189 i \cr 0.3168 + 0.189 i & -0.2667 + 0.3075 i \cr
\end {pmatrix} \qquad
 P= \begin{pmatrix}-0.236 - 0.2001 i &
0.1787 - 0.028 i \cr 0.1787 - 0.028 i    & -0.2297 + 0.1202 i\cr \end {pmatrix} \nnu
A&=& \begin{pmatrix} -0.23 - 0.3521 i & 0.3382 + 0.0777 i \cr 0.3382 + 0.0777 i
&-0.4475 + 0.2227 i \cr \end {pmatrix} \quad
\tilde\sigma=\tilde{\bar\sigma}=\begin{pmatrix}
 0.0863 - 0.2366 i &  0.1973 + 0.1041 i \cr
 0.1973 + 0.1041 i  &
   -0.0863 + 0.2366 i
\cr \end {pmatrix} \nnu \bar D_X &=& 2(|p_+|^2-|p_-|^2
+3(|a_+|^2-|a_-|^2)+6(|w_+|^2-|w_-|^2) +
{\frac{1}{2}}|\sigma_+|^2-|\sigma_-|^2 +
{\frac{1}{2}}|\bar{\sigma}_+|^2-|\bar{\sigma}_-|^2)\nonumber\\&& =-8.94\eea

\subsection{$N_g$=3}
As discussed in \cite{BMsugry} traceless part of BM field $S_{ab}$
is used to cancel $O(N_g)$ D-terms. For fixing $\hat S_{ab}$,
without loss of generality,  one can rotate the fields by an
$O(3)$ transformation so that the
 so that only  one component of the  D-term vector is non zero. VEVs
below are written in the basis where only the third D-term
component    $<D^3>$ is non zero. This basis is denoted  by a
prime on all VEVs.
\subsubsection{ $Rank[\mathbf{\Xi}]=$5 (tracefull symmetric representation)}

\bea \tilde\sigma'=\tilde{\bar\sigma}' &=& \begin{pmatrix} -8.1532 + 14.4793 i & -11.7404 - 7.4196 i&
  0.194 - 1.1043 i \cr -11.7404 - 7.4196 i & 6.6912 - 9.4853 i &
  0.9135 + 0.2088 i \cr 0.194 - 1.1043 i & 0.9135 + 0.2088 i &
  0.0124 + 0.0746 i\end{pmatrix}\nnu
A'&=&\begin{pmatrix}-2.7937 - 0.4459 i & -0.0298 - 1.8182 i &
  2.573 - 0.2572 i\cr -0.0298 - 1.8182 i & -2.6651 +
   0.5903 i & -0.4297 - 1.5744 i \cr 2.573 - 0.2572 i & -0.4297 -
   1.5744 i & -0.8325 - 0.1193 i\end{pmatrix}\eea
\bea P'&=&\begin{pmatrix}2.6802 - 0.9855 i &
  0.589 - 1.518 i& -13.9797 + 0.005 i \cr 0.589 - 1.518 i &
  4.1178 + 0.5751 i& 1.9362 - 0.8664 i\cr -13.9797 + 0.005 i&
  1.9362 - 0.8664 i& 6.1195 + 0.1825 i\end{pmatrix}\nnu
W'&=&\begin{pmatrix}1.0342 + 0.1527 i & -0.4614 + 0.2617 i &
  2.0891 + 0.4169 i \cr -0.4614 + 0.2617 i & -1.4776 -
   0.1941 i& -0.3529 + 2.8359 i \cr 2.0891 + 0.4169 i & -0.3529 +
   2.8359 i & 1.3112 + 0.0711 i\end{pmatrix}\eea
\be {D'}_{X}^a = 3319.0 \delta^a_3  \ee

\subsubsection{ $Rank[\mathbf{\Xi}]=$4 (tracefull symmetric representation) }

\bea \tilde\sigma'=\tilde{\bar\sigma}'  &=& \begin{pmatrix}-0.0515 - 2.2441 i & 1.6389 + 0.9556 i &
  0.8735 + 1.689 i \cr 1.6389 + 0.9556 i &
  0.4307 + 0.2916 i & -1.8341 - 2.1534 i \cr 0.8735 +
   1.689 i & -1.8341 - 2.1534 i & 0.926 + 1.8694 i
  \end{pmatrix}\nnu
A' &=& \begin{pmatrix} -1.1162 + 0.1493 i & -0.6261 - 0.4133 i & -0.1611 +
  0.4422 i \cr -0.6261 - 0.4133 i &
 0.3311 + 0.7292 i & -0.3128 - 0.1764 i \cr -0.1611 +
  0.4422 i & -0.3128 - 0.1764 i & -0.7771 - 0.6807 i
 \end{pmatrix} \eea
\bea P' &=&\begin{pmatrix} 0.7956 - 0.2474 i & 0.0406 - 0.4419 i &
 0.6712 + 0.2745 i \cr 0.0406 - 0.4419 i &
 0.3486 + 1.7078 i & -0.0373 - 0.1994 i \cr 0.6712 +
  0.2745 i & -0.0373 - 0.1994 i & -0.4144 - 0.7517 i
 \end{pmatrix} \nnu
W' &=&\begin{pmatrix} -0.0025 - 0.4259 i &
 0.0644 + 0.4904 i & -0.3268 - 0.5708 i \cr 0.0644 +
  0.4904 i & -0.3575 - 0.0647 i &
 0.0554 + 0.2052 i \cr -0.3268 - 0.5708 i & 0.0554 + 0.2052 i &
 0.5644 + 0.6209 i \end{pmatrix} \eea
\be {D'}_{X}^a = 47.04 \delta^a_3  \ee

\subsection{ $Rank[\mathbf{\Xi}]=$4 (traceless symmetric
representation) }

\bea
\tilde\sigma'=\tilde{\bar\sigma}' &=&\begin{pmatrix} 1.5506 - 5.4394 i & 1.7398 + 4.1109 i &
 0.0011 - 0.0811 i \cr 1.7398 + 4.1109 i & -1.6868 + 2.0758 i &
 0.0202 + 0.042 i \cr 0.0011 - 0.0811 i & 0.0202 + 0.042 i &
 0.1361 + 3.3636 i \end{pmatrix} \nnu
 A'&=&\begin{pmatrix}-1.8304 - 0.7199 i &
 1.5213 - 0.445 i & -0.0854 + 0.0056 i \cr 1.5213 - 0.445 i &
 0.6854 + 0.7288 i & -0.0906 - 0.0529 i \cr -0.0854 +
  0.0056 i & -0.0906 - 0.0529 i & 1.145 - 0.009 i
  \end{pmatrix} \nnu
 P'&=&\begin{pmatrix}-0.2976 + 0.0331 i &
 0.3609 - 0.0673 i & -0.2904 + 0.0343 i \cr 0.3609 - 0.0673 i &
 0.2394 + 0.4181 i & -0.5147 - 0.2562 i \cr -0.2904 +
  0.0343 i &  -0.5147 - 0.2562 i & 0.0581 - 0.4512 i
 \end{pmatrix}  \nnu
 W'&=&\begin{pmatrix} 1.0622 + 0.3626 i & -0.7826 + 0.3279 i & -0.0567 +
  0.0102 i \cr -0.7826 + 0.3279 i & -0.3667 - 0.2597 i & -0.1406 -
  0.0597 i \cr -0.0567 + 0.0102 i & -0.1406 - 0.0597 i & -0.6955 -
  0.1029 i \end{pmatrix}  \eea
\be {D'}_{X}^a = 275.15 \delta^a_3  \ee
\section*{\bf {APPENDIX B : ILLUSTRATIVE TABLES OF YUKAWA STRUCTURE AND SUPERHEAVY SPECTRA IN YUMGUTs}}
In this Appendix, we give the Yukawa hierarchies and complete
superheavy spectra for the  three example solutions discussed
above, in units of the MSGUT scale parameter
$m/\lambda$. This   is  done  to remove any prejudice that
unacceptable  pseudo-Goldstone multiplets must occur in family
symmetric models. The naming convention  and MSGUT multiplicities
can be found in\cite{ag2}.  \pagestyle{empty}
\begin{table}
 $$
 \begin{array}{|c|c|c|c|c|c|}
 \hline
   {\rm  S.No.  }&  {\rm M_H}& {\rm  Y_u}&  {\rm Y_d} &{\rm \theta_{CKM}(deg.)}&m_\nu(meV)\\
 \hline
 1.&0.049 + 0.190 i& 0.1537, 0.0080& 0.0537,
  0.0043& 5.15&  0.0255,  0.2791 \\
2.&0.599 + 0.791 i& 0.1293, 0.0118& 0.0562,
  0.0051& 2.27 \times 10^{-6}&0.0013, 0.0144
\\
3.&1.39 + 0.80 i& 0.0685, 0.0214& 0.0359, 0.0052& 7.41&0.0011,  0.0121 \\
\hline
   {\rm  S.No.  }&{\rm Y_l}&Y_{\nu}&{\rm \theta_{PMNS}(deg.)}&\tilde{M}_{\nu^c}&\Delta m_\nu^2(eV^2) \\
 \hline
1.& 0.0424, 0.0027& 0.2515,
  0.0233& 14.5& 0.6967, 0.0636& 7.73 \times 10^{-8} \\
 2.&  0.0712, 0.0065& 0.0576,
  0.0053& 2.32 \times 10^{-5}& 0.6969, 0.0636& 2.06 \times 10^{-10} \\
3.&0.0147,
  0.0063& 0.0911, 0.0028& 33.7& 0.6969, 0.0636& 1.45 \times 10^{-10} \\ \hline
  \end{array}$$
  \caption{ Yukawa eigenvalues and mixing angles for $N_g=2, f=-0.13$ . $\tilde{M}_{\nu^c}\equiv
   \lambda  {M}_{\nu^c}/m$. $m/\lambda $ is taken to be $10^{16}$ GeV
  to estimate $\Delta m_\nu^2 $. $\lambda = -0.038 + .005 i$, $\eta = 0.4$, $\gamma = 0.32$, $\bar\gamma = -1.6$, $h = .34$, $\xi=0.8719+.5474i$.
  $M_{\nu^c}$ is independent of $M_H$ value chosen. \label{table:ng=2} }
 \end{table}
 \begin{table}
 $$
 \begin{array}{|c|c|c|c|c|c|}
 \hline
   {\rm  S.No.  }&  {\rm M_H}& {\rm  Y_u}&  {\rm Y_d} &{ \rm \theta_{CKM}(deg.)}&m_\nu(meV)\\
 \hline
1.&0.049 + 0.190 i& 0.1761, 0.0131& 0.0507,
  0.0038& 0.00486&  10.05,  110.19\\
2.&0.599 + 0.791 i& 0.1108,0.0101& 0.0569,
  0.0052& 2.47 \times 10^{-9}&  4.86, 53.29\\
3.&  1.39 + 0.80 i& 0.0721, 0.0140& 0.0283,
  0.00552& 0.00767& 4.39,  48.12\\
\hline
   {\rm  S.No.  }&{\rm Y_l}&Y_{\nu}&{\rm \theta_{PMNS}(deg.)}&\tilde{M}_{\nu^c}&\Delta m_\nu^2(eV^2) \\
 \hline
1.& 0.0507, 0.0038& 0.1762,
  0.0131& 8.7& 0.000697, 0.0000636& 0.01204\\
 2.&  0.0569,
  0.0052& 0.1108, 0.0101& 2.79 \times 10^{-6}& 0.000697,
  0.0000636& 0.00282\\
 3.&  0.0283, 0.0055& 0.0721,
  0.0140& 26.8& 0.000697, 0.0000636& 0.00230\\
  \hline \end{array}$$
   \caption{ Effect of reducing $f$ :
  Yukawa eigenvalues and mixing angles for $N_g=2, f=-0.00013$ and other parameters same as in Table
  \ref{table:ng=2}. Notice that the   light neutrino masses are in an  acceptable range
  but  $Y_u=Y_\nu,Y_d=Y_l$ and the   quark mixing is negligible.\label{table:ng=2fsmall} }
 \end{table}

\begin{table}[!htp]
 $$
 \begin{array}{|c|c|c|c|}
 \hline
 {\rm  Field }&\hspace{10mm} {\rm Masses}& {\rm  Field }&\hspace{6mm} {\rm Masses}\\
 {\rm [SU(3),SU(2),Y]}& &{\rm [SU(3),SU(2),Y]} & \\ \hline
       A[1,1,4]&  {4.093, 3.321, 0.137}& B[6,2,{5/3}]&   {0.106, 0.099, 0.091}      \\
          C[8,2,1]&{1.727, 1.727, 1.224}& C[8,2,1]&  1.224, 0.614, 0.614\\
           D[3,2,{7/ 3}]&  1.919, 1.433, 1.191  &D[3,2,{7/ 3}]& 0.810, 0.205, 0.134  \\
  E[3,2,{1/3}]& 1.475, 1.043, 0.716  &E[3,2,{1/3}]&  0.716, 0.677, 0.594,0.506\\
  E[3,2,{1/3}] &{{0.404, 0.277, 0.087}}& E[3,2,{1/3}] &  0.073, 0.050, 0.004 \\
 F[1,1,2]&   1.794, 1.794, 1.681 & F[1,1,2]& 1.317, 0.289, 0.228, 0.018\\
  G[1,1,0]&{1.672, 1.665, 1.248}& G[1,1,0]& 1.248, 0.766, 0.766, 0.504 \\
G[1,1,0]&  0.469, 0.208, 0.201&G[1,1,0]& 0.079, 0.068, 0.055\\
G[1,1,0]&0.011, 0.009,{\color{red} 0} &h^{(1)}[1,2,1]& {3.799, 2.812, 1.398, 1.182, 0.983 }\\
h^{(1)}[1,2,1]& { 0.74,0.588, 0.511, 0.159, 0.024, 0.013} &h^{{(2)}}[1,2,1]&{ 3.947, 2.961, 1.623, 1.247, 1.009 } \\
h^{{(2)}}[1,2,1]& { 0.726, 0.556, 0.51, 0.14, 0.044,0.005} & h^{{(3)}}[1,2,1]&{ 4.161, 3.196, 2.049}\\
 h^{{(3)}}[1,2,1]&{ 1.289, 0.979, 0.710}&h^{{(3)}}[1,2,1]&  {.540, .520, .152,  .029,  .010 }\\
  I[3,1,{10/3}]&  {0.210, 0.192, 0.003}& J[3,1,{4/3}]& 1.889, 1.889, 0.946  \\
 J[3,1,{4/3}]&{{  0.740, 0.453, 0.278  }}&J[3,1,{4/3}]&{ 0.119,0.086, 0.021, 0.006 }\\
 K[3,1, {-8/ 3}]&{ 1.591, 1.237, 0.116} &L[6,1,{2/ 3}]&{ 1.066, 0.916, 0.757 } \\
  M[6,1,{8/ 3}]& {1.340, 0.958, 0.493} &  N[6,1,{-4/ 3}]& {1.795, 1.178, 0.345}    \\
   O[1,3,-2]& {1.127, 0.886, 0.084} & P[3,3,{-2/ 3}]&{ {0.902, 0.754,
   0.595}}\\
    Q[8,3,0]& {0.163, 0.126, 0.083}&  R[8,1, 0]&  0.170, 0.119, 0.107\\
             R[8,1, 0]&{  0.086, 0.066, 0.047} &  S[1,3,0]& {0.090, 0.058, 0.011} \\
      t^{{(1)}}[3,1,{-2/ 3}]&{ 3.264, 2.802, 1.824, 1.496, 1.175 }  &  t^{{(1)}}[3,1,{-2/ 3}]& { 1.019, 0.89, 0.824, 0.598, 0.495}  \\
             t^{{(1)}}[3,1,{-2/ 3}]&{ 0.343, 0.202, 0.055, 0.026, 0.007} &  t^{{(2)}}[3,1,{-2/ 3}]&{ 3.418, 2.936, 1.873, 1.636, 1.2} \\
              t^{{(2)}}[3,1,{-2/ 3}]&{ 1.053, 0.909, 0.824, 0.692, 0.532} &  t^{{(2)}}[3,1,{-2/ 3}]&{ 0.454, 0.211, 0.077, 0.018, 0.001}  \\
  t^{{(3)}}[3,1,{-2/ 3}]&{ 3.650, 3.156, 2.097,1.747, 1.273}& t^{{(3)}}[3,1,{-2/ 3}]& {  1.116,0.926,0.824, 0.779,0.541}\\
    t^{{(3)}}[3,1,{-2/ 3}]&  {{0.466, 0.223,0.116, 0.023,0.002} } & U[3,3,{4/3}]& {0.084, 0.070, 0.054} \\
       V[1,2,-3]& {0.227, 0.208, 0.003} & W[6,3,{2/ 3}]&    {1.693, 1.324, 0.902}   \\
  X[3,2,{-5/ 3}]&  1.666, 1.666, 0.149&  X[3,2,{-5/ 3}]&   .102,  .072,  .070,  .066\\
  Y[6,2, {-1/3}]&   {0.167, 0.118, 0.058}&  Z[8,1,2]&   {0.100, 0.086, 0.070}        \\
  \hline
 \end{array}
 $$
  \caption{ Mass Spectrum of superheavy fields in units of $m/\lambda \sim 10^{16}$ GeV for $N_g$=2
   for the solution described in Section IV.A. Only the spectra of $h[1,2,\pm 1],t[3,1,\mp2/3]$ depend on the three possible values of
      $M_H$ (see Table I for the values of $M_H$). The corresponding sets are labelled as $h^{(i)},t^{(i)},i=1,2,3.$
   {\label{Ngeq2PGoldies}}}\end{table}
\begin{table}
 $$
 \begin{array}{|c|c|c|c|c|}
 \hline
   {\rm  S.No.  }&  {\rm M_H}& {\rm  Y_u}&  {\rm Y_d} &{\rm \{\theta_{13}, \theta_{12}, \theta_{23}\}^Q }\\
   &&&&(deg.)  \\
 \hline
1.&2.55 + 0.13 i &0.007, 0.019, 0.368 &0.007, 0.014, 0.306 &0.56, 13.18, 1.58 \\
2.&1.44 - 0.61 i &0.027, 0.13, 0.409 &0.009, 0.083, 0.242 &3.42, 8.71, 3.87 \\
3.&1.28 + 0.75 i &0.063, 0.228, 0.424 &0.019, 0.083, 0.186 &6.65, 6.59, 1.11 \\
4.&1.16 + 0.67 i & 0.062, 0.193, 0.439 & 0.02, 0.099, 0.188 &2.6, 5.18, 1.96 \\
5.&1.06 - 0.73 i & 0.009, 0.076, 0.458 &0.008, 0.078, 0.321  &1.51, 11.19, 4.61 \\
6.&0.02 - 0.03 i &0.022, 0.254, 0.604 & 0.009, 0.104, 0.289 & 1.04, 1.57, 6.03\\
 \hline
   {\rm  S.No.  }&{\rm Y_l}&Y_{\nu}&m_\nu(eV)&{M}_{\nu^c} \\
 \hline
1.&0.007, 0.026, 0.421 &0.014, 0.032, 0.533 & 0.16, 0.1597, 0.0104, 0.0104,1.7 \times 10^{-6}   &365.07,0,0 \\
2.&0.023, 0.094, 0.314 &0.018, 0.213, 0.566& 0.2056, 0.2054, 0.05, 0.0499, 3.8 \times 10^{-6} & 365.07,0,0 \\
3.&0.031, 0.103, 0.212 &0.015, 0.187, 0.624 &0.3021, 0.302, 0.0781, 0.0781, 4.4 \times 10^{-7} & 365.07,0,0 \\
4.&0.029, 0.094, 0.259 &0.018, 0.283, 0.42 &0.1806, 0.1805, 0.1254, 0.1254, 7.8\times 10^{-7} & 365.07,0,0 \\
5.&0.009, 0.073, 0.4 &0.008, 0.214, 0.558 &0.1946, 0.1945, 0.0533, 0.0532 ,7.2\times 10^{-7}  & 365.07,0,0 \\
6.&0.007, 0.148, 0.338 &0.01, 0.159, 0.608 &0.2837, 0.2836,
0.0129, 0.0128, 6.4\times 10^{-6}  & 365.07,0,0 \\\hline
   \end{array}$$
   \caption{ Yukawa eigenvalues and mixing angles for $N_g=3$($Rank[\mathbf{\Xi}]=$5), $f=0.9+0.7 i$. $\lambda  =.48+.3i $, $\eta =.25  $,
    $h = 1.3$, $\gamma =.05  $, $\bar\gamma  =-1.2$, $\xi =3.645+.363i $. $M_{\nu^c}$ is independent of $M_H$ value chosen.
    Notice the four neutrino masses in the $10^{1\div 2}$ meV range.\label{table:rank5}}
 \end{table}
\clearpage

\begin{table}[!htp]
 $$
 \begin{array}{|c|c|c|c|}
 \hline
 {\rm  Field }&\hspace{10mm} {\rm Masses}& {\rm  Field }&\hspace{6mm} {\rm Masses}\\
 {\rm [SU(3),SU(2),Y]}& &{\rm [SU(3),SU(2),Y]} & \\ \hline
       A[1,1,4]&  {9.93, 8.08, 7.34, 4.06, 2.53, 0.84}& B[6,2,{5/3}]&   {6.46, 6.38, 5.8, 5.18, 4.83, 2.88}      \\
          C[8,2,1]&{4.62, 4.62, 4.36, 4.36, 4., 4.}& C[8,2,1]& 2.2, 2.2, 1.88, 1.88, 0.35, 0.35 \\
           D[3,2,{7/ 3}]& {7.55, 4.64, 4.25, 3.95, 2.71, 2.} &D[3,2,{7/ 3}]& 1.92, 1.61, 1.39, 1.1, 1.08, 0.53 \\
  E[3,2,{1/3}]& {  28.48, 28.48, 20.62, 19.48, 18.45 }  &E[3,2,{1/3}]&  18.19,14.85, 13.38, 5.94, 4.47, \
3.32
  \\
  E[3,2,{1/3}] &{2.87  , 2.54, 2.28, 2.2, 1.54, 1.36, 1.28 }& E[3,2,{1/3}] & 1.16,1.06, 0.97, 0.8,
0.4, 0.25, 0.24 \\
 F[1,1,2]& 24.66, 24.66, 21.78, 20.8, 19.4, 18.4 & F[1,1,2]& 14.65, 13.88, 4.6, 3.65,
1.9, 1.7, 0.14 \\
  G[1,1,0]&{45.55, 45.55, 39.11, 37.34, 36.03

  }& G[1,1,0]& 35.82,27.46, 26.75, 15.96, 14.2,14.0 \\
G[1,1,0]&  13.83,11.92, 11.41, 10.75, 8.62 &G[1,1,0]& 7.15 ,6.35, 6.03, 5.27, 4.47, 3.4, 2.15\\
G[1,1,0]& 1.77, 0.76, 0.66, 0.11, 0.05,{\color{red} 0, 0, 0}&
  I[3,1,{10/3}]&  {17.23, 13.25, 11.82, 4.61, 2.61, 1.17}\\ J[3,1,{4/3}]& 25.09, 25.09, 23.7, 23.02, 20.65, 19.5&
  J[3,1,{4/3}]&{16.49, 15.45, 7.68, 7.07,
5.18,4.66 }\\J[3,1,{4/3}]&{ 4.51, 3.25, 3.07, 1.94, 0.86, 0.68, 0.23}&
 K[3,1, {-8/ 3}]&{8.41, 5.21, 4.54, 1.46, 1.01, 0.44 } \\L[6,1,{2/ 3}]&{{13.27, 9.28, 5.3, 4.92, 3.49, 0.91}} &
  M[6,1,{8/ 3}]& {{12.48, 8.46, 6.42, 4.37, 3.29, 0.71}} \\  N[6,1,{-4/ 3}]& {{14.23, 10.71, 7.19, 7.07, 3.87, 0.22}}    &
   O[1,3,-2]& {{14.05, 9.35, 8.34, 5.48, 3.22, 1.05}}\\ P[3,3,{-2/ 3}]&{9.63, 7.14, 5.65, 2.8, 1.6, 1.27}&
    Q[8,3,0]& {15.14, 12.38, 8.53, 7.75, 2.31, 0.93}\\  R[8,1, 0]&{27.29, 18.82, 13.31, 10.7, 9.3, 6.36} &
             R[8,1, 0]&{6.12, 6.07, 5.45, 3.68, 2.58, 0.15} \\  S[1,3,0]& {29.76, 19.87, 14.44, 10.53, 7.96, 1.74} &
              U[3,3,{4/3}]& {{24.78, 15.76, 13.17, 6.8, 5.83, 3.13}}\\
       V[1,2,-3]& {18.19, 16.8, 15.59, 13.91, 4.48, 2.43} &W[6,3,{2/ 3}]&    {{6.72, 5.37, 4.41, 2.72, 1.44, 0.15}}  \\
  X[3,2,{-5/ 3}]&16.48, 16.48, 11.8, 9.05, 8.37, 6.22 &  X[3,2,{-5/ 3}]&  3.69, 3.5, 2.79,1.91, 1.31,
0.77, 0.49  \\
Y[6,2, {-1/3}]&   {{7.51, 7., 6.1, 3.69, 2.2, 1.52}}&  Z[8,1,2]&   {{27.05, 18.02, 10.94, 9.02, 8.06, 0.93}}        \\
  \hline
 \end{array}
 $$
  \caption{Mass Spectrum of superheavy fields in units of $m/\lambda \sim 10^{16} $ GeV in
 six-dimensional symmetric tracefull($Rank[\mathbf{\Xi}]=5$) scenario with  $N_g$=3 for the solution described in Section IV.B.1.
 Only the spectra of $h[1,2,\pm 1],t[3,1,\mp2/3]$ depend on the value of
      $M_H$ chosen. See Table \ref{htrank5-6dim} for the  spectra of $h[1,2,\pm 1],t[3,1,\mp2/3]$ type multiplets for
       each of the six values of $M_H$.\label{tracefullrank5spec}}\end{table}
\clearpage
\begin{table}[!htp]
 $$
 \begin{array}{|c|c|c|}
 \hline {\mbox {$M_H$ } }&h[1,2,1] &t[3,1,-2/3]\\
 \hline
2.55 + 0.13 i& 34.81, 32.15, 27.12, 21.08, 20.53 & 39., 36.27, 35.17, 27.86, 27.6, 24.06, 21.67 \\

&18.99,12.6, 12.06, 11.62, 10.96,
10.38& 18.22, 13.95, 12.81,10.72, 8.94, 8.28, 7.46, 6.5\\
& 9.51, 7.29 , 6.87, 3.57, 2.7,2.28, 2.06 & 5.07, 4.49, 4.09, 3.44, 3.08, 2.23,1.52, 1.36\\
&  0.74,0.59, 0.32,0.13, 0.05 & 0.78, 0.49, 0.43, 0.26, 0.21,
0.11, 0.04 \\\hline
1.44 - 0.61 i& 34.77, 32.11, 27.08, 21.06, 20.5, 18.95 &38.97, 36.25, 35.14, 27.88, 27.49, 24.01, 21.68\\
&12.5, 11.96, 11.55, 10.81, 10.28, 9.54  &18.09, 13.86, 12.87, 10.71, 8.73, 8.17, 7.54 \\
& 7.3, 6.87, 3.5, 2.59, 2.31, 1.97 &6.53, 5.05, 4.49, 4.07, 3.45, 2.92, 2.24, 1.49 \\
& 0.5, 0.38, 0.23, 0.11, 0.04 & 1.29, 0.61, 0.44, 0.37, 0.23,
0.18, 0.1, 0.05\\\hline
1.28 + 0.75 i&34.77, 32.1, 27.08, 21.05, 20.5, 18.95  &38.97, 36.25, 35.13, 27.88, 27.48, 24., 21.69, 18.08\\
&12.49, 11.95, 11.55, 10.81, 10.27, 9.54 &13.86, 12.87, 10.7, 8.72, 8.16, 7.55, 6.53, 5.05 \\
&7.3, 6.86, 3.5, 2.59, 2.32, 1.96 & 4.49, 4.07, 3.45, 2.91, 2.24, 1.49, 1.29\\
& 0.49, 0.37, 0.22, 0.11, 0.01 &0.58, 0.44, 0.35, 0.21, 0.17, 0.1, 0.06 \\
\hline
1.16 + 0.67 i& 34.77, 32.1, 27.07, 21.05, 20.49, 18.94 &38.96, 36.24, 35.13, 27.88, 27.47, 24., 21.69, 18.07\\
&12.48, 11.94, 11.54, 10.79, 10.26, 9.55 &13.85, 12.88, 10.7, 8.7, 8.15, 7.56, 6.54, 5.05 \\
& 7.3, 6.86, 3.49, 2.58, 2.32, 1.95 & 4.49, 4.07, 3.46, 2.89, 2.24, 1.49, 1.28\\
&0.45, 0.34, 0.21, 0.12, 0.01 &0.55, 0.43, 0.33, 0.21, 0.16, 0.1, 0.06 \\
\hline
1.06 - 0.73 i &34.77, 32.1, 27.07, 21.05, 20.49, 18.94  &38.96, 36.24, 35.13, 27.89, 27.46, 24., 21.69, 18.07\\
&12.48, 11.94, 11.54, 10.79, 10.25, 9.55 & 13.85, 12.88, 10.7, 8.69, 8.15, 7.56, 6.54, 5.05\\
& 7.31, 6.86, 3.49, 2.58, 2.32, 1.95 & 4.49, 4.07, 3.45, 2.89, 2.24, 1.49, 1.28\\
& 0.46, 0.34, 0.22, 0.14, 0.03& 0.54, 0.43, 0.34, 0.21, 0.16, 0.1, 0.06\\
\hline
0.02 - 0.03 i & 34.75, 32.08, 27.05, 21.04, 20.47, 18.92 &38.95, 36.24, 35.12, 27.91, 27.4, 23.97, 21.69, 18.\\
&12.44, 11.9, 11.51, 10.74, 10.19, 9.57 &13.82, 12.92, 10.68, 8.58, 8.1, 7.6, 6.56, 5.05 \\
&7.33, 6.86, 3.48, 2.57, 2.33, 1.9 & 4.5, 4.06, 3.46, 2.84, 2.24, 1.48, 1.25\\
&0.36, 0.28, 0.23, 0.21, 0.09 &0.43, 0.31, 0.22, 0.19, 0.13, 0.11, 0.06 \\
 \hline
 \end{array}
 $$
 \caption{\small{Mass Spectrum of superheavy fields $h[1,2,\pm 1],t[3,1,\mp2/3]$ which depend on the value of
      $M_H$ chosen in units of $m/\lambda \sim 10^{16} $ GeV in
 six-dimensional symmetric tracefull($Rank[\mathbf{\Xi}]= 5$) scenario with  $N_g$=3 for each of the solutions described in Section IV.B.1.\label{htrank5-6dim}}}\end{table}

\begin{table}
 $$
 \begin{array}{|c|c|c|c|c|c|}
 \hline
   {\rm  S.No.  }&  {\rm M_H}& {\rm  Y_u}&  {\rm Y_d} &{\rm \{\theta_{13}, \theta_{12}, \theta_{23}\}^Q }&m_\nu/10^{-4}(meV)\\
   &&&&(deg.) & \\
 \hline
1.&-4.323+1.47i&.0007,.0021,.0215&.001,.0019,.0219&9.0,15.9,15.6&0.039 ,0.187,40.543 \\
2.&.465+3.382i&.0018,.0148,.0182&.0020,.0197,.0222&11.3,1.5,4.9&0.079 ,0.807 ,4.09 \\
3.&.76-2.193i&.0029,.0113,.0137&.0054,.0233,.0385&1.5,6.2,7.4&0.12 ,0.32 ,6.946 \\
4.&-0.002+0.968i&.0105,.040,.077&.0035,.0174,.0408&5.2,3.7,2.8&1.33 ,4.503 ,23.439 \\
5.&-.508-.209i&.0077,.053,.1126&.0019,.0159,.0381&1.1,12.1,1.4&0.926 ,17.046,34.448\\
6.&-.092-.032i&.0041,.0467,.0558&.0035,.0413,.0522&8.7,5.5,2.6&1.03,4.96,9.806\\
 \hline
   {\rm  S.No.  }&{\rm Y_l}&Y_{\nu}&{\rm \{\theta_{13}, \theta_{12}, \theta_{23}\}^L }&\tilde{M}_{\nu^c}&\Delta m_\nu^2/10^{-13}(eV^2) \\
   &&&(deg.)& & \\
 \hline
1.&.0013,.0041,.0517&.0023,.0064,.0468&3.6,20.3,23.3&5.9,5.3,1.5&.0033,164.372\\
2.&.0034,.0148,.0205&.0032,.0126,.0162&27.5,14.5,47.0&5.9,5.3,1.5&0.0645,1.6074\\
3.&.0053,.0121,.0458&.0033,.0102,.020&13.6,11.1,41.1&5.9,5.3,1.5&0.0088,4.8143\\
4.&.0048,.0174,.0473&.0092,.0181,.0915&23.9,14.5,17.7&5.9,5.3,1.5&1.851,52.9128\\
5.&.0042,.0224,.0382&.0061,.0584,.0835&23.7,26.1,49.4&5.9,5.3,1.5&28.9722,89.6078\\
6.&.0043,.0497,.0621&.0049,.0355,.0518&14.1,37.6,46.3&5.9,5.3,1.5&2.3544,7.1562\\\hline
   \end{array}$$
   \caption{ Yukawa eigenvalues and mixing angles for $N_g=3 (Rank[\mathbf\Xi]=4)$,
 $f= -0.11  + .02 i$. $\tilde{M}_{\nu^c}\equiv
   \lambda  {M}_{\nu^c}/m$. $m/\lambda $ is taken to be $10^{16}$ GeV
  to estimate $\Delta m_\nu^2 $. $\lambda  = 0.48 - .05  i$, $\eta = -.18$, $h = .26$, $\gamma = 0.12$, $\bar\gamma  =
  -1.44$ and $\xi = 1.7278  - 0.1734i $\label{table:tracefulldata}}
 \end{table}

\begin{table}
 $$
 \begin{array}{|c|c|c|c|c|c|}
 \hline
   {\rm  S.No.  }&  {\rm M_H}& {\rm  Y_u}&  {\rm Y_d} &\rm {\{\theta_{13}, \theta_{12}, \theta_{23}\}^Q}&m_\nu(meV)\\
   &&&& (deg.)&\\
 \hline
1.&-4.323+1.47i&0.001,0.003,0.026&0.001,0.003,0.027&0.002,0.008,0.015&0.0006,0.0047,1.2427\\
2.&.465+3.382i&0.002,0.013,0.015&0.002,0.015,0.017&0.014,0.002,0.003&0.0056,0.0284,0.3286\\
3.&.76-2.193i&0.003,0.01,0.014&0.005,0.019,0.028&0.002,0.008,0.013&0.0096,0.0256,0.3686\\
4.&-0.002+0.968i&0.007,0.033,0.079&0.004,0.017,0.042&0.005,0.004,0.002&0.0652,0.8982,3.3757\\
5.&-.508-.209i&0.006,0.047,0.105&0.002,0.017,0.038&0.001,0.013,0.002&0.0701,2.2717,3.2128\\
6.&-.092-.032i&0.003,0.044,0.054&0.003,0.043,0.053&0.009,0.006,0.004&0.025,0.7852,1.4609\\
\hline
  {\rm  S.No.  }&{\rm Y_l}&Y_{\nu}&{\rm \{\theta_{13}, \theta_{12}, \theta_{23}\}^L}&\tilde{M}_{\nu^c}&\Delta m_\nu^2/10^{-5} (eV^2) \\
  &&& (deg.)&&\\
   \hline
1.&0.001,0.003,0.027&0.001,0.003,0.026&0.27,9.82,2.13&0.006,0.005,0.001&2.16\times 10^{-6},0.15\\
2.&0.002,0.015,0.017&0.002,0.013,0.015&2.41,22.56,37.83&0.006,0.005,0.001&7.74\times 10^{-5},0.011\\
3.&0.005,0.019,0.028&0.003,0.01,0.014&2.97,28.08,18.45&0.006,0.005,0.001&5.64 \times 10^{-5},0.013 \\
4.&0.004,0.017,0.042&0.007,0.033,0.079&5.81,7.17,25.52&0.006,0.005,0.001&0.080 ,1.06 \\
5.&0.002,0.017,0.038&0.006,0.047,0.105&2.7,6.33,54.28&0.006,0.005,0.001&0.516 ,0.516 \\
6.&0.003,0.043,0.053&0.003,0.044,0.054&3.38,7.5,58.18&0.006,0.005,0.001&0.062,0.152\\\hline
 \end{array}$$
   \caption{(Truncated) Yukawa eigenvalues and mixing angles for $N_g=3$ and the same parameter values as in Table \ref{table:tracefulldata} but
  with much smaller $f= -0.00011  + .00002 i$. $\tilde{M}_{\nu^c}\equiv
   \lambda  {M}_{\nu^c}/m$. $m/\lambda $ is taken to be $10^{16}$ GeV
  to estimate $\Delta m_\nu^2 $.\label{table:tracefulldata-fsmall} }
 \end{table}
\clearpage
\begin{table}[!htp]
 $$
 \begin{array}{|c|c|c|c|}
 \hline
 {\rm  Field }&\hspace{10mm} {\rm Masses}& {\rm  Field }&\hspace{6mm} {\rm Masses}\\
 {\rm [SU(3),SU(2),Y]}& &{\rm [SU(3),SU(2),Y]} & \\ \hline
       A[1,1,4]&  {{1.53, 1.46, 1.28, 1.26, 0.38, 0.07}}& B[6,2,{5/3}]&   {{2.89, 2.44, 2.41, 1.9, 1.22, 1.}}      \\
          C[8,2,1]&{{1.2, 1.2, 0.91, 0.91, 0.89, 0.89}}& C[8,2,1]& 0.63, 0.63, 0.6, 0.6, 0.59, 0.59 \\
           D[3,2,{7/ 3}]& {1.27, 0.9, 0.86, 0.72, 0.67, 0.63} &D[3,2,{7/ 3}]& 0.61, 0.51, 0.41, 0.35, 0.19, \
0.13 \\
  E[3,2,{1/3}]& {6.26, 6.26, 4.38, 3.93, 3.54, 2.99}  &E[3,2,{1/3}]&
  2.94, 2.51, 1.44, 1.12, 1.09, \
1.02 \\
  E[3,2,{1/3}] &{0.97, 0.92, 0.77, 0.73, 0.7, 0.68, 0.6}& E[3,2,{1/3}] &  0.42, 0.32, 0.24, 0.22, \
0.19, 0.12 \\
 F[1,1,2]& 6.01, 6.01, 5.12, 3.97, 3.44, 3.18,2.95 & F[1,1,2]&  2.77, 1.42, 1.04, 0.46, \
0.19, 0.12\\
  G[1,1,0]&{10.19, 10.19, 8.23, 7.9, 6.58, 6.27}& G[1,1,0]& 5.52, 5., 3.99, 3.75, 2.88, \
2.79\\
G[1,1,0]& 2.53, 2.22, 2.15, 2.03, 1.79, 1.71 &G[1,1,0]& 1.67, 1.33, 0.97, 0.95,0.88, 0.79
\\
G[1,1,0]& 0.62, 0.42, 0.06,{\color{red} 0, 0, 0, 0}&
  I[3,1,{10/3}]&  {2.89, 2.74, 2.42, 2.37, 1.46, 0.3}\\ J[3,1,{4/3}]& 6.4, 6.4, 4.68, 3.93, 3.45, 3.09  &
  J[3,1,{4/3}]&{ 2.97, 2.64, 1.68, 1.27, 1.19, 1.14}\\J[3,1,{4/3}]&{
0.96, 0.9, 0.54, 0.37, 0.16, 0.11, 0.03  }&
 K[3,1, {-8/ 3}]&{{1.01, 0.77, 0.76, 0.63, 0.56, 0.47} } \\L[6,1,{2/ 3}]&{1.53, 1.08, 1.04, 0.63, 0.61, 0.6} &
  M[6,1,{8/ 3}]& {{1.58, 1.33, 1.27, 1.05, 0.78, 0.39}} \\  N[6,1,{-4/ 3}]& {{1.63, 1.31, 0.98, 0.93, 0.7, 0.49}}    &
   O[1,3,-2]& {{1.36, 1.22, 0.81, 0.7, 0.49, 0.22}}\\ P[3,3,{-2/ 3}]&{{0.8, 0.7, 0.62, 0.36, 0.19, 0.18}}&
    Q[8,3,0]& {{2.89, 2.69, 2.09, 1.59, 1.26, 1.24}}\\  R[8,1, 0]&{3.54, 2.89, 2.66, 2.52, 2.43, 1.9} &
             R[8,1, 0]&{1.78, 1.53, 1.32, 0.89, 0.88, \
0.77} \\  S[1,3,0]& {{2.92, 1.89, 1.51, 0.89, 0.84, 0.51}} &
             U[3,3,{4/3}]& {{1.52, 1.48, 1.17, 0.96, 0.28, 0.26}} \\
       V[1,2,-3]& {{3.77, 2.59, 1.96, 0.96, 0.86, 0.77}} & W[6,3,{2/ 3}]&    {{1.27, 1.1, 1., 0.83, 0.73, 0.7}}   \\
  X[3,2,{-5/ 3}]& 2.89, 2.33, 2.33, 2.08, 1.8, 1.75 &  X[3,2,{-5/ 3}]& 1.37, 1.06, 0.96, 0.92,0.88, \
0.87, 0.34  \\
Y[6,2, {-1/3}]&   {{2.51, 1.74, 1.68, 0.97, 0.91, 0.89}}&  Z[8,1,2]&   {{3.39, 2.18, 2.12, 1.16, 0.98, 0.96}}        \\
  \hline
 \end{array}
 $$
 \caption{Mass Spectrum of superheavy fields in units of
  $m/\lambda \sim 10^{16} $ GeV in six-dimensional symmetric tracefull scenario($Rank[\mathbf{\Xi}]=4$)  for  $N_g=3 $
  for the solution described in Section IV.B.2. Only the spectra of $h[1,2,\pm 1],t[3,1,\mp2/3]$ depend on the value of
      $M_H$ chosen.  See Table \ref{htrank4-6dim} for these  spectra.\label{tracefullspec} }\end{table}

\clearpage
 \begin{table}[!htp]
 $$
 \begin{array}{|c|c|c|}
 \hline {\mbox {$M_H$ } }&h[1,2,1] &t[3,1,-2/3]\\
 \hline
-4.323+1.47i&10.09, 9.86, 8.22, 7.93, 7.31, 6.46 &9.96, 9.82, 7.84, 7.49, 7.12, 6.57, 5.94 \\

&5.85, 3.74, 3.49, 3.26, 2.39, 2.04&3.43, 3.29, 3.05, 2.22, 1.84, 1.72, 1.49, 1.32, 1.16 \\
&1.76, 1.41, 0.89, 0.75, 0.63, 0.61 &1.05, 1., 0.99, 0.79, 0.74, 0.65, 0.63, 0.61 \\
&  0.48, 0.33, 0.27, 0.1, 0.03 & 0.52, 0.4, 0.32, 0.28, 0.2,
0.12\\\hline
.465+3.382i&9.71, 9.48, 7.76, 7.42, 6.71, 6.27, 5.16 & 9.63, 9.39, 7.35, 7.03, 6.59, 6.28, 5.11\\

&3.58, 3.17, 2.64, 2.4, 1.96, 1.72& 3.18, 2.91, 2.7, 2.1, 1.83,1.68, 1.46, 1.36\\
&1.55, 0.92, 0.66, 0.63, 0.61 & 1.21, 1.11, 1.04, 1.02, 0.8, 0.74, 0.66, 0.63\\
&0.52, 0.32, 0.26, 0.13, 0.03  &  0.61, 0.49, 0.37, 0.25, 0.22,
0.17, 0.13\\\hline
.76-2.193i&9.52, 9.22, 7.52, 7.1, 6.35, 5.94 &9.46, 9.09, 7.14, 6.74, 6.23, 5.87, 4.46, 3.01 \\

&4.61, 3.29, 2.87, 2.52, 2.09, 1.92&2.51, 2.14, 2.09, 1.91, 1.72, 1.43, 1.28, 1.2 \\
&1.76, 1.17, 0.97, 0.72, 0.66, 0.62 &1.12, 1.07, 1.03, 0.79, 0.77, 0.63, 0.61 \\
&  0.52, 0.31, 0.22, 0.17, 0.04 &0.59, 0.47, 0.35, 0.27, 0.18,
0.14, 0.1 \\\hline

-0.002+0.968i&9.4, 9.01, 7.38, 6.9, 6.16, 5.7 &9.34, 8.87, 7.04, 6.54, 5.98, 5.61, 3.84, 2.83 \\

&4.19, 3.18, 2.61, 2.36, 2.11, 1.38& 2.25, 1.99, 1.83, 1.69, 1.64, 1.38, 1.32, 1.29\\
&1.15, 1.07, 1.01, 0.73, 0.66, 0.63 & 1.17, 1.06, 0.95, 0.82, 0.79, 0.63, 0.61\\
& 0.36, 0.2, 0.13, 0.06, 0.02 &0.57, 0.38, 0.2, 0.13, 0.1, 0.08,
0.03 \\\hline
-.508-.209i&9.39, 8.99, 7.37, 6.86, 6.15, 5.64 & 9.33, 8.84, 7.03, 6.5, 5.96, 5.54, 3.76, 2.84\\
&4.12, 3.16, 2.45, 2.41, 2.13, 1.35& 2.1, 1.99, 1.79, 1.7, 1.62, 1.46, 1.35, 1.27\\
&1.24, 0.95, 0.82, 0.71, 0.67, 0.63 & 1.14, 0.98, 0.91, 0.82, 0.81, 0.63, 0.59 \\
&  0.39, 0.17, 0.11, 0.05, 0.02 &0.55, 0.28, 0.14, 0.12, 0.08,
0.06, 0.03 \\\hline
-.092-.032i&9.37, 8.98, 7.35, 6.84, 6.13, 5.61 & 9.32, 8.82, 7.03, 6.48, 5.95, 5.52, 3.71, 2.83\\

&4.1, 3.17, 2.45, 2.37, 2.17, 1.32&2.07, 1.99, 1.82, 1.7, 1.62, 1.45, 1.33 \\
&1.2, 0.88, 0.85, 0.72, 0.68, 0.63 &1.27, 1.14, 1., 0.89, 0.83, 0.81, 0.63, 0.59 \\
& 0.37, 0.12, 0.08, 0.05, 0.02
  &0.55, 0.31, 0.1, 0.07, 0.06, 0.04, 0.02 \\\hline
 \end{array}
 $$
 \caption{\small{Mass Spectrum of superheavy fields $h[1,2,\pm 1],t[3,1,\mp2/3]$ which depend on the value of
      $M_H$ chosen in units of $m/\lambda \sim 10^{16} $ GeV in
 six-dimensional symmetric tracefull($Rank[\mathbf{\Xi}]= 4$) scenario with  $N_g$=3 for each of the solutions described in Section IV.B.2.
 \label{htrank4-6dim}}}\end{table}

\begin{table}
 $$
 \begin{array}{|c|c|c|c|c|c|}
 \hline
   {\rm  S.No.  }&  {\rm M_H}& {\rm  Y_u}&  {\rm Y_d} &\rm {\{\theta_{13}, \theta_{12}, \theta_{23}\}^Q}&m_\nu(meV)\\
     &&&& (deg.)&\\
 \hline
1.&135.29+11.98  i&.054,.068,.1718& .0001,.0012,.0019&8.23,34.35,32.04&(1.29 ,1.62 ,4.1)\times 10^{-2}\\
2.&25.4+1.72  i& 0.0,.0588,.0938&0.0,.0021,.0034&(.1,1.7,1.2)\times 10^{-6}&6.5\times 10^{-11},.012,.028\\
3.&24.6+1.16  i& 0.0,.0614,.1325&0.0,.0016,.0036&8.12\times 10^{-8},0,0&2.0\times 10^{-10},.017,.045\\
4.&18.41+1.4  i&.001,.0555,.125&.0035,.011,.0264&3.48,6.01,8.67&1.5\times 10^{-6},.010,.011\\
5.&18.32+1.23  i&0.0,.0584,.1322&0.0,.0112,.0255&4.86\times
10^{-9},0,0&1.1\times 10^{-10},.012,.013\\ \hline
  {\rm  S.No.  }&{\rm Y_l}&Y_{\nu}&\rm {\{\theta_{13}, \theta_{12}, \theta_{23}\}^L}&\tilde{M}_{\nu^c}&\Delta m_\nu^2(eV^2) \\
    &&& (deg.)&&\\
 \hline
1.&.0029,.0066,.013&.174,.2183,.551&5.92,32.15,9.6&22.39,8.89,7.05&(.096,1.42)\times 10^{-9}\\
2.&0.0,.0177,.0283&0.0,.2272,.3629&(.11,8.5)\times 10^{-7},28.7&22.39,8.89,7.05&(1.49,6.18)\times 10^{-10}\\
3.&0.0,.0068,.0147& 0.0,.236,.5094&(.03,1.2)\times 10^{-6},19.4&22.39,8.89,7.05&(2.75,17.9)\times 10^{-10}\\
4.&.0098,.0136,.038&.0028,.122,.275&14.33,20.53,35.13&22.39,8.89,7.05&(1.08,.23)\times 10^{-10}\\
5.&0.0,.015,.034& 0.0,.1284,.291&5.7\times 10^{-9},0,42.03&22.39,8.89,7.05&(1.33,.29)\times 10^{-10}\\
  \hline
   \end{array}$$
   \caption{ Yukawa eigenvalues and mixing angles for $N_g=3 (Rank[\mathbf\Xi]=4)$,
 $f= 0.23 +.04 i$ . $\tilde{M}_{\nu^c}\equiv
   \lambda  {M}_{\nu^c}/m$. $m/\lambda $ is taken to be $10^{16}$ GeV
  to estimate $\Delta m_\nu^2 $. $\lambda = 0.18 - .03i$,
$\eta$ = .034, $\gamma$ = -0.53 $, \bar\gamma$ = -2.60, $h = .14$
and $\xi= 7.677  + 0.15772i$. $M_{\nu^c}$ is independent of $M_H$
value chosen. \label{table:tracelessdata}}  \end{table}

\clearpage
\begin{table}[!htp]
 $$
 \begin{array}{|c|c|c|c|}
 \hline
 {\rm  Field }&\hspace{10mm} {\rm Masses}& {\rm  Field }&\hspace{6mm} {\rm Masses}\\
 {\rm [SU(3),SU(2),Y]}& &{\rm [SU(3),SU(2),Y]} & \\ \hline
       A[1,1,4]&  {{0.602, 0.596, 0.515, 0.461, 0.446}}& B[6,2,{5/3}]&   {{1.099, 0.861, 0.826, 0.248, 0.244}}      \\
          C[8,2,1]&{0.563, 0.563, 0.552, 0.552, 0.532 }& C[8,2,1]& 0.532, 0.483, 0.483, 0.482, 0.482 \\
           D[3,2,{7/ 3}]& 0.584, 0.571, 0.563, 0.561, 0.531  &D[3,2,{7/ 3}]& 0.513, 0.493, 0.482, 0.465, 0.461 \\
  E[3,2,{1/3}]& 8.843, 8.843, 1.48, 1.47, 0.92   &E[3,2,{1/3}]&0.858, 0.847, 0.658, 0.658 , 0.482\
 \\
  E[3,2,{1/3}] &{0.467,0.462, 0.452, 0.426, 0.423}& E[3,2,{1/3}] &  0.388, 0.351, 0.33, 0.31, 0.061, 0.022  \\
 F[1,1,2]& 8.262, 8.262, 1.603, 1.591, 0.997 & F[1,1,2]& 0.901  , 0.882, 0.49, 0.462, 0.165, 0.075 \\
  G[1,1,0]&{15.3, 15.3, 2.62, 2.601, 1.866}& G[1,1,0]&   1.317,1.315, 1.111, 1.11,0.979
  \\
G[1,1,0]&0.92, 0.772 , 0.744, 0.642, 0.626 &G[1,1,0]& 0.498, 0.384,0.327, 0.327, 0.245
\\
G[1,1,0]&0.136, 0.065, 0.031, {\color{red}0, 0, 0 }&
  I[3,1,{10/3}]&  {0.974, 0.898, 0.462, 0.149, 0.06}\\ J[3,1,{4/3}]& 9.188, 9.188, 1.438, 1.428, 0.908  &
  J[3,1,{4/3}]&{0.83, 0.821, 0.573, 0.525,0.494,0.469}\\J[3,1,{4/3}]&{ 0.432, 0.21, 0.19, 0.07, 0.025 }&
 K[3,1, {-8/ 3}]&{{0.565, 0.555, 0.53, 0.483, 0.48 } } \\L[6,1,{2/ 3}]&0.596, 0.574, 0.542, 0.452, 0.45 &
  M[6,1,{8/ 3}]& {{0.692, 0.635, 0.582, 0.359, 0.356}} \\  N[6,1,{-4/ 3}]& {{0.549, 0.548, 0.516, 0.503, 0.496}}    &
   O[1,3,-2]& {{0.772, 0.771, 0.419, 0.396, 0.282}}\\ P[3,3,{-2/ 3}]&{{0.596, 0.593, 0.504, 0.471, 0.45}}&
    Q[8,3,0]& {{0.967, 0.846, 0.595, 0.14, 0.112}}\\  R[8,1, 0]& 1.18, 0.959, 0.856, 0.441, 0.406  &
             R[8,1, 0]&{0.394, 0.319, 0.31, 0.274, 0.272 } \\  S[1,3,0]& {1.279, 1.269, 0.537, 0.098, 0.024} &
              U[3,3,{4/3}]& {0.776, 0.762, 0.3, 0.138, 0.051} \\
       V[1,2,-3]& {1.065, 1.05, 0.422, 0.167, 0.076} & W[6,3,{2/ 3}]&    {0.63, 0.602, 0.549, 0.421, 0.416}   \\
  X[3,2,{-5/ 3}]& 2.084, 2.084, 0.857, 0.791,0.772 &  X[3,2,{-5/ 3}]&  0.699,0.577, 0.238, 0.16, 0.07, 0.053 \\
Y[6,2, {-1/3}]&   {0.587, 0.533, 0.426, 0.167, 0.158}&  Z[8,1,2]&   {0.768, 0.669, 0.501, 0.063, 0.045}        \\
  \hline
 \end{array}
 $$
 \caption{Mass Spectrum of superheavy fields in units of
 $m/\lambda \sim 10^{16} $ GeV in the five-dimensional symmetric traceless case($N_g=3, Rank[\mathbf\Xi]=4)$,
 for the solution described
 in Section IV.C. Only the spectra of $h[1,2,\pm 1],t[3,1,\mp2/3]$ depend on the value of
      $M_H$ chosen. See Table\ref{htrank4-5dim} for these  spectra. \label{tracelessspec}}\end{table}

\clearpage
 \begin{table}[!htp]
 $$
 \begin{array}{|c|c|c|}
 \hline {\mbox {$M_H$ } }&h[1,2,1] &t[3,1,-2/3]\\
 \hline
135.29+11.98  i&136.697, 136.68, 136.528, 135.959&136.517, 136.501, 136.392, 135.942, 135.923, 2.293 \\
& 135.936,2.939,2.925, 1.921,0.78, 0.77 & 2.282, 1.407,0.697, 0.687, 0.652, 0.609, 0.593 \\
&0.655, 0.62,0.557,0.55, 0.505, 0.481  &0.573, 0.559, 0.527, 0.522, 0.52, 0.483 \\
&  0.466, 0.266,0.217 &0.477, 0.391, 0.387, 0.169, 0.139, 0.072
\\\hline

25.4+1.72  i&29.744, 29.664, 28.968, 26.179, 26.063&28.908, 28.83, 28.323, 26.094, 25.996, 2.715 \\
&3.287, 3.252, 1.821, 1.52, &2.688, 1.506, 1.195, 0.86, 0.809, 0.679 \\
&0.992, 0.927, 0.658, 0.605, 0.579 & 0.633, 0.599, 0.572, 0.556, 0.534, 0.526, 0.398\\
&0.517, 0.284, 0.115, 0.103, 0.008  &0.392, 0.258, 0.177, 0.161,
0.026, 0.007 \\\hline

24.6+1.16  i&29.034, 28.952, 28.237, 25.37, 25.25&28.176, 28.096, 27.575, 25.282, 25.181, 2.732 \\
&3.299, 3.263, 1.828, 1.55 &2.703, 1.533, 1.202, 0.868, 0.816, 0.682 \\
&1.003, 0.937, 0.659, 0.606, 0.585 &0.635, 0.599, 0.572, 0.556, 0.534, 0.526, 0.398 \\
&0.518, 0.285, 0.104, 0.092, 0.008  &0.392, 0.259, 0.17, 0.153,
0.031, 0.012 \\\hline

18.41+1.4  i&24.01, 23.909, 23.03, 19.435, 19.281&22.965, 22.865, 22.217, 19.321, 19.19, 2.89 \\
&3.413, 3.353, 2.055, 1.669 &2.841, 1.831, 1.228, 0.942, 0.885, 0.734, 0.644 \\
&1.115, 1.043, 0.697, 0.633, 0.602 &0.604, 0.577, 0.562, 0.537, 0.528, 0.398 \\
&0.528, 0.291, 0.076, 0.061, 0.004  & 0.392, 0.276, 0.109, 0.095,
0.069, 0.061\\\hline

18.32+1.23  i&23.926, 23.825, 22.943, 19.332, 19.177& 22.878, 22.777, 22.127, 19.218, 19.086\\
&3.416, 3.355, 2.062, 1.669 & 2.893, 2.844, 1.838, 1.228, 0.943, 0.887, 0.735\\
&1.118, 1.046, 0.699, 0.633, 0.602 &  0.645, 0.604, 0.577, 0.562, 0.537, 0.528, 0.398\\
& 0.528, 0.291, 0.078, 0.062, 0.004  & 0.392, 0.276, 0.108, 0.095,
0.069, 0.061\\\hline

 \end{array}
 $$
 \caption{\small{Mass Spectrum of superheavy fields $h[1,2,\pm 1],t[3,1,\mp2/3]$ which depend on the value of
      $M_H$ chosen in units of $m/\lambda \sim 10^{16} $ GeV in
 five-dimensional symmetric traceless($Rank[\mathbf{\Xi}]=4$) scenario with  $N_g$=3 for each of the solutions described in Section IV.C.\label{htrank4-5dim}}}\end{table}

\clearpage

\section*{Acknowledgements} C.S.A. is  grateful to Borut Bajc for
 numerous valuable discussions. C.K.K. thanks
the University Grants Commission of the Government of India for
financial support through an  UGC-Senior research fellowship.


\begin{thebibliography}{99}



\bibitem{RparitySO10}C.S.~Aulakh, B.~Bajc, A.~Melfo, A.~Rasin and G.~Senjanovic,
  Nucl.\ Phys.\ B {\bf 597}, 89 (2001)
  [arXiv:hep-ph/0004031].


\bibitem{aulmoh} C.S.~Aulakh and R.N.~Mohapatra, CCNY-HEP-82-4 April 1982,
  CCNY-HEP-82-4-REV,  Jun 1982 , Phys. Rev. {\bf D28}, 217 (1983).


\bibitem{ckn} T.E.~Clark, T.K.~Kuo, and N.~Nakagawa, Phys. lett. {\bf{115B}}, 26(1982).

\bibitem{abmsv}
C.S.~Aulakh, B.~Bajc, A.~Melfo, G.~Senjanovic and F.~Vissani,
Phys.\ Lett.\ B {\bf 588}, 196 (2004) [arXiv:hep-ph/0306242].


\bibitem{blmdm}C.S.~Aulakh, \emph{From germ to bloom},
 arXiv:hep-ph/0506291 ; C.S.~Aulakh and S.K.~Garg,
  Nucl.\ Phys.\ B {\bf 757}, 47 (2006)
  [arXiv:hep-ph/0512224].

\bibitem{nmsgut}C.S.~Aulakh and S.K.~Garg, Nucl.
  Phys.\textbf{B}857 (2012)101, arXiv:0807.0917v3.


 \bibitem{bstabhedge}C.S.~Aulakh,
  arXiv:hep-ph/1107.2963 ;
  C.S.~Aulakh, I.~Garg and C.K.~Khosa,
 Nucl.\ Phys.\ B {\bf 882}, 397(2014), arXiv:1311.6100 [hep-ph].


\bibitem{spurion} 
  N.~Cabibbo and L.~Maiani,
  Evolution of particle physics, pages 50-80 (1970). For a useful recent pedagogical review see R.A. de Pablo,
arXiv:1307.1904v1[hep-ph].


\bibitem{koide}
  Y.~Koide,
  Phys.\ Rev.\ D {\bf 78}, 093006 (2008)
  [arXiv:0809.2449 [hep-ph]];  
  Phys.\ Rev.\ D {\bf 79}, 033009 (2009)
  [arXiv:0811.3470 [hep-ph]];
  Phys.\ Lett.\ B {\bf 665}, 227 (2008).

\bibitem{ag2}
  C.S.~Aulakh and A.~Girdhar,
    Nucl.\ Phys.\ B {\bf 711}, 275 (2005).

\bibitem{ag1} C.S.~Aulakh and A.~Girdhar,
   Int.\ J.\ Mod.\ Phys.\ A {\bf 20}, 865 (2005)

\bibitem{bmsv}B.~Bajc, A.~Melfo, G.~Senjanovic and F.~Vissani,
Phys.\ Rev.\ D {\bf 70}, 035007 (2004) [arXiv:hep-ph/0402122];


 \bibitem{BMsugry} C.S.~Aulakh, `` \emph{Bajc-Melfo Vacua enable YUMGUTs}'',
arXiv:1402.3979

 \bibitem{ray}
  S.~Ray,
  Phys.\ Lett.\  B {\bf 642} (2006) 137
  [arXiv:hep-th/0607172].


 \bibitem{BM}
  B.~Bajc and A.~Melfo,
  JHEP {\bf 0804}, 062 (2008)
  [arXiv:0801.4349 [hep-ph]].


\bibitem{witten}
  E.~Witten,
  Phys.\ Lett.\  B {\bf 105} (1981) 267.


 \bibitem{ovrab}
  B.A.~Ovrut and S.~Raby,
  Phys.\ Lett.\ B {\bf 125}, 270 (1983).

\bibitem{hilo}
  C.S.~Aulakh,
   ``Local Supersymmetry And Hi-lo Scale Induction,''
  CCNY-HEP-83/2;  C.S.~Aulakh,
 PhD Thesis, City University of New York, 1983, UMI-84-01477.


 \bibitem{damalibra}
  R.~Bernabei, P.~Belli, S.~d'Angelo, A.~Di Marco, F.~Montecchia, F.~Cappella, A.~d'Angelo and A.~Incicchitti {\it et al.},
  Int.\ J.\ Mod.\ Phys.\ A {\bf 28}, 1330022 (2013)
  [arXiv:1306.1411 [astro-ph.GA]].



 \bibitem{modulicon}  G.D.~Coughlan, W.~Fischler, E.W.~Kolb, S.~Raby and G.~G.~Ross,
  Phys.\ Lett.\ B {\bf 131} (1983) 59; B.~de Carlos, J.~A.~Casas, F.~Quevedo and E.~Roulet,
  Phys.\ Lett.\ B {\bf 318} (1993) 447
  [hep-ph/9308325].


 \bibitem{realcore} C.S.~Aulakh,
  hep-ph/0602132.

 \bibitem{bertschwmal} S.~Bertolini, T.~Schwetz and M.~Malinsky,
  Phys.\ Rev.\ D {\bf 73} (2006) 115012
  [hep-ph/0605006].


\bibitem{bmsvdecem} B.~Bajc, A.~Melfo, G.~Senjanovic and F.~Vissani,
  Phys.\ Lett.\ B {\bf 634}, 272 (2006)
  [hep-ph/0511352].





\bibitem{fuku04} T.~Fukuyama, A.~Ilakovac, T.~Kikuchi, S.~Meljanac and N.~Okada,
  Eur.\ Phys.\ J.\ C {\bf 42}, 191 (2005)
  arXiv:hep-ph/0401213v1.,v2.

\end{thebibliography}
\end{document}